\documentclass[final,3p,times]{elsarticle}

\usepackage{graphicx}
\usepackage{amssymb}
\usepackage{amsthm}
\usepackage{amsmath}
\usepackage{xcolor}
\usepackage[linesnumbered,ruled]{algorithm2e} 

\DeclareMathOperator*{\argmin}{arg\!\min}
\allowdisplaybreaks
\usepackage{slashbox}
\usepackage{lineno}
\usepackage{float}
\usepackage{array,multirow,graphicx}

\usepackage[usestackEOL]{stackengine}
\usepackage{graphicx}
\stackMath
\def\dashfill{\cleaders\hbox to .5em{\rule{.4ex}{.4pt}}\hfill}
\newcommand\dashline[1]{\hbox to #1{\dashfill\hfil}}
\newlength\tmplen
\usepackage{arydshln}

\makeatletter
\renewcommand*\env@matrix[1][\arraystretch]{%
  \edef\arraystretch{#1}%
  \hskip -\arraycolsep
  \let\@ifnextchar\new@ifnextchar
  \array{*\c@MaxMatrixCols c}}
\makeatother

\makeatletter
\DeclareRobustCommand{\abs}{\@ifstar\star@abs\normal@abs}
\newcommand{\normal@abs}[2][]{\mathopen{#1|}#2\mathclose{#1|}}
\makeatother

\usepackage{array}
\newcommand{\PreserveBackslash}[1]{\let\temp=\\#1\let\\=\temp}
\newcolumntype{C}[1]{>{\PreserveBackslash\centering}p{#1}}
\newcolumntype{R}[1]{>{\PreserveBackslash\raggedleft}p{#1}}
\newcolumntype{L}[1]{>{\PreserveBackslash\raggedright}p{#1}}

\newtheorem{theorem}{Theorem}

\newtheorem{lemma}[theorem]{Lemma}
\newtheorem{pro}{Proposition}

\theoremstyle{definition}

\theoremstyle{remark}

\theoremstyle{remark}
\newtheorem{problem}{Problem}[section]
\theoremstyle{remark}
\newtheorem{exmp}{Example}[section]

\usepackage{appendix}

\journal{Comput. Methods Appl. Mech. Engrg.}

\begin{document}

\begin{frontmatter}


\title{FEA-Net: A Physics-guided Data-driven Model for Efficient Mechanical Response Prediction}

\author{Houpu Yao, Yi Gao, Yongming Liu*}
\address{Arizona State University, Tempe, Arizona, 85281, United States}

\begin{abstract}

An innovative physics-guided learning algorithm for predicting the mechanical response of materials and structures is proposed in this paper. The key concept of the proposed study is based on the fact that physics models are governed by Partial Differential Equation (PDE), and its loading/ response mapping can be solved using Finite Element Analysis (FEA). Based on this, a special type of deep convolutional neural network (DCNN) is proposed that takes advantage of our prior knowledge in physics to build data-driven models whose architectures are of physics meaning. This type of network is named as FEA-Net and is used to solve the mechanical response under external loading. Thus, the identification of a mechanical system parameters and the computation of its responses are treated as the learning and inference of FEA-Net, respectively. Case studies on multi-physics (e.g., coupled mechanical-thermal analysis) and multi-phase problems (e.g., composite materials with random micro-structures) are used to demonstrate and verify the theoretical and computational advantages of the proposed method. 

\end{abstract}

\begin{keyword}
Physics-guided  Learning \sep Data-driven Model \sep Convolutional Neural Networks \sep Finite Element Analysis  


\end{keyword}

\end{frontmatter}



\section{Introduction}
Predicting physics is important for various real-world applications. For example, in remaining life prediction of mechanical systems \cite{liu2006fatigue}, weather forecasting \cite{maqsood2004ensemble}, and earthquake alert \cite{dutta2002geopressure}. Both data-driven and physics-based solutions have been developed for fast and reliable predictions, yet both approaches have their own limitations. In the following, we briefly review these approaches and their limitations, before proposing a solution which integrates the two entirely different methodologies.

Recent successes of deep learning for computer vision, speech recognition, natural language processing, and control~\cite{krizhevsky2012imagenet,amodei2016deepspeech,devlin2018bert,silver2017mastering} have inspired studies on data-driven approaches to prediction tasks in engineering contexts. For example, some of the seminal studies have investigated the application of deep learning in thermal \cite{sheikholeslami2019nanofluid} and fluid \cite{tompson2016fluid,chu2017smoke} simulations, structure analysis \cite{wang2019load_path,finol2018eigen} and optimization \cite{sosnovik2017neural,cang2018cad}, material property prediction \cite{bouman2013estimating,li2019predicting_shale} and design \cite{bessa2017framework,cang2018improving}, system monitoring \cite{zhao2019DL_SHM} and calibration \cite{yao2019sensorcalibration}. These tasks either lack physics-based models or have models that are expensive to compute, rendering data-driven approaches reasonable alternatives.

Despite the empirical success of machine learning models, deep neural networks (DNNs) in particular, the following key challenges still remain yet are often overlooked: (1) \emph{Lack of generalizability}: Generalizability measures how well a model learns on finite samples of a data distribution performs on other samples from the same distribution. Recent studies have exposed the lack of generalizability of machine learning models~\cite{wang2019FSL}. In the meantime, the acquisition of large physics dataset can be expensive either experimentally or computationally ~\cite{cang2018improving}, which makes the application of data-driven model to physics less efficient.
(2) \emph{Lack of interpretability}: DNN is often criticized for their lack of interpretability in engineering applications. In engineering contexts, this means that physically meaningful insights cannot be generated from the observations. Furthermore, users of data-driven models often have little knowledge about how and why the models may fail.

On the other hand, physics-based models, such as Finite Element Analysis (FEA), have long been developed to model mechanical systems. 
Mechanical systems are governed by basic physical principles such as conservation laws or minimum energy, which can be expressed into Partial Differential Equations (PDE). Such physics-based models can be very accurate and fully interpretable, but the constitutive law of the system needs to be explicitly obtained first. One drawback of the resultant physics-based models is that they are problem dependent, which means different models needs to be built for different problems. 
In addition, such models usually have relatively larger computational cost for large systems. 

\begin{figure} 
\label{fig:key_idea}
  \centering
    \includegraphics[width=0.8\linewidth]{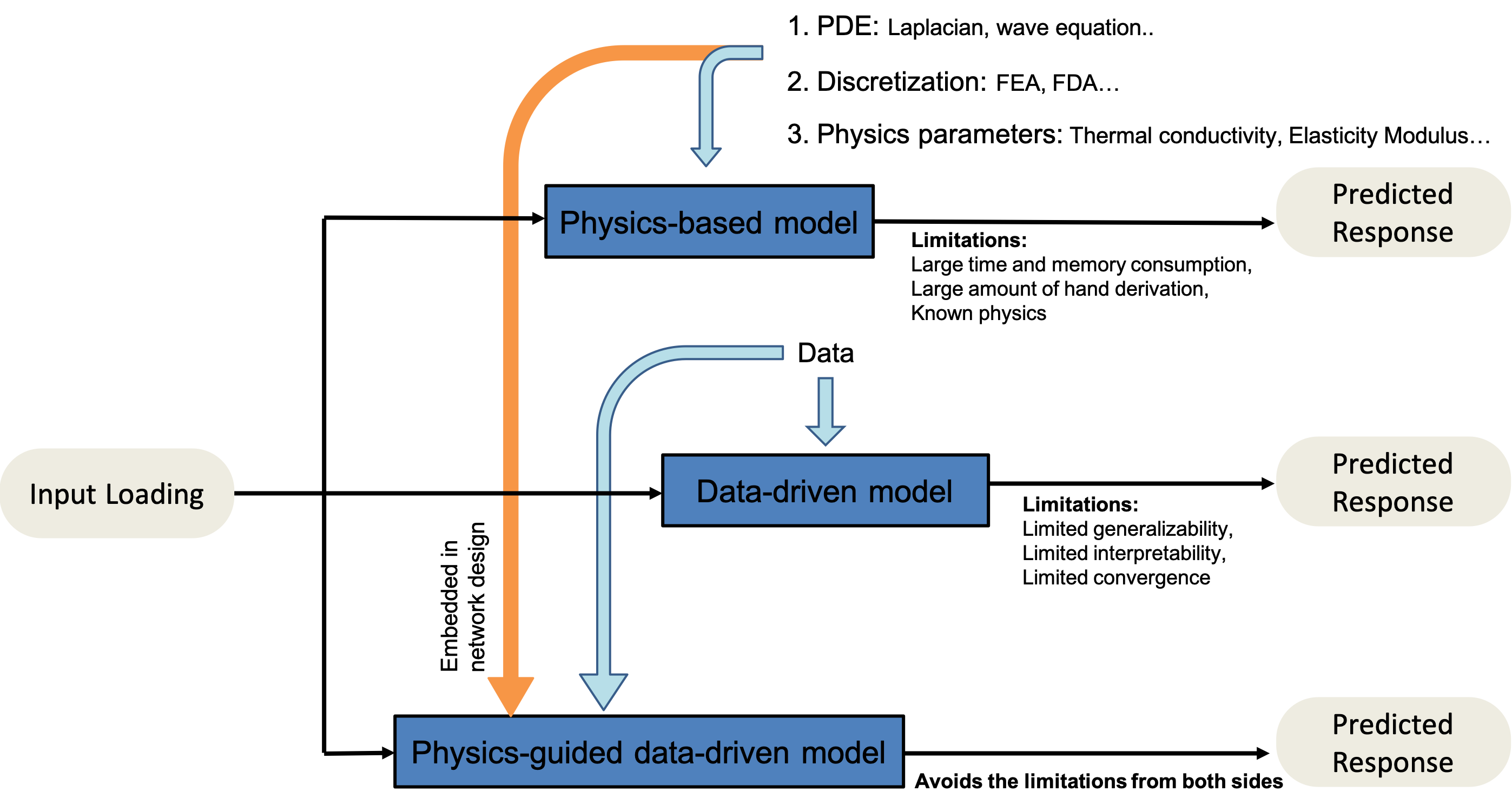}
    \caption{Illustration of a FEA-Net. Yellow arrow means this knowledge is embedded in the network structure by design.
    }
\end{figure}

Due to above-mentioned challenges for both methodologies, it is appealing to develop a hybrid model that leverages the generalizability of physics-based models and flexibility of data-driven models. 
As illustrated in Fig.~\ref{fig:key_idea}, our prior knowledge in physics principles is used to guide the designing of neural network structures.
\emph{The key insight} of this paper is that physics problems are governed by PDEs, and FEA models for PDEs are actually a special type of deep convolutional neural networks (which we coin as ``FEA-Net''). Therefore, the identification and the computation of responses of a mechanical system can be reformulated as network learning and inference tasks, respectively. FEA-Nets can be easily implemented in matured learning infrastructures (e.g., TensorFlow~\cite{abadi2016tensorflow} and pytorch~\cite{paszke2017pytorch}) to leverage GPU computation. 

An outline of the paper is as follows:
We introduce notations and the problem statement in the rest of this section. Sec.\ref{sec:background} reviews related work from deep learning and computational mechanics. 
Sec.\ref{sec:fea_conv} introduces FEA convolution, an operator that critically enables the connection between FEA models and DNNs. We then extend FEA convolution to handle multi-physics and multi-phase PDEs in Sec.\ref{sec:fea_net}.
Verification studies through numerical examples are discussed in Sec.\ref{sec:experiment}. 
Sec.\ref{sec:conclusion} concludes the paper.


\section{Preliminaries and background}
\label{sec:background}

\subsection{Problem statement}
\label{subsec:problem_def}

To start with, we make the following {assumptions}: (1) All loading/ response observations are in image form. This assumption makes the use of convolutional neural networks as data-driven model possible. (2) Consider linear physics only.  As with \cite{kawaguchi201no_local_minima,hsieh2018ICLR_PDE}, we start with simpler linear physics first since it is easier to prove the convergence of the proposed algorithm. Future work will extend this framework to non-linear physics and irregular mesh data. 

\begin{figure}%
\label{fig:heatmap}
  \centering
    \includegraphics[width=1.0\linewidth]{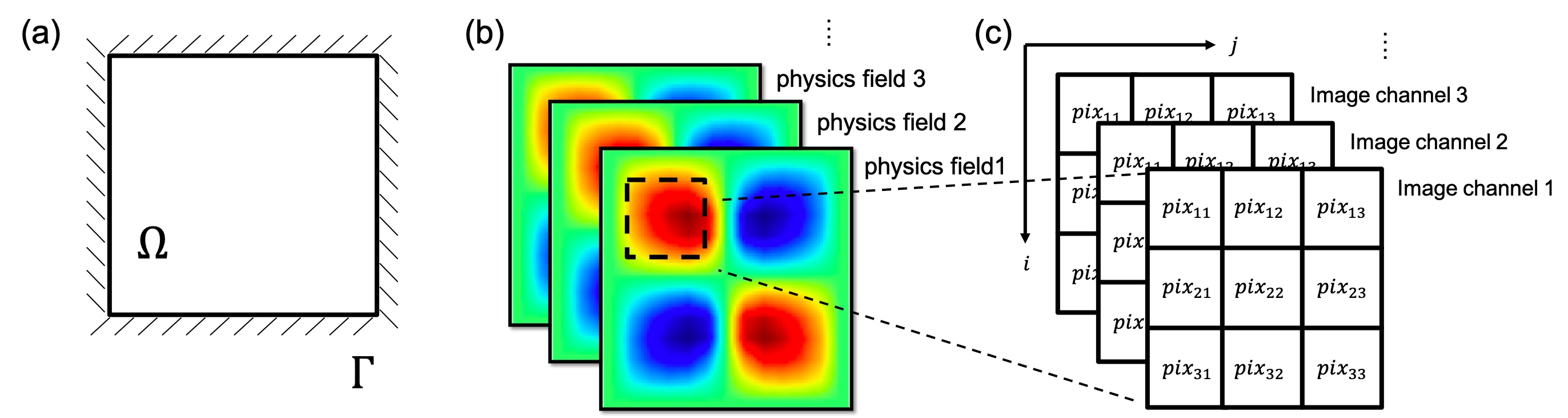}
    \caption{
    (a) Solution domain $\Omega$ with boundary condition.
    (b) Load/ response heatmap can be viewed as images.
    (c) Image pixels and their coordinate system.
    }
\end{figure}

As an example of the first assumption, consider a solution domain $\Omega$ is 2D and square-shaped as depicted in Fig.~\ref{fig:heatmap}a. There can be multiple different physics fields in $\Omega$, which can be visualized as several different heatmaps (Fig.~\ref{fig:heatmap}b). These heatmaps can be viewed as a multiple channel image, as shown in Fig.~\ref{fig:heatmap}c. For example, there are three channels for thermoelasticity problems: x- and y- directional displacement (or force) and temperature (or heat flux). In the rest of this paper, the loading and response images are denoted as $V \in R^{(N,N,P)}$ and $U \in R^{(N,N,Q)}$ respectively, where $N$ is the spatial resolution of the images, $P$ and $Q$ are the number of input and output channels for loading and response images respectively. 

We consider a dataset ${\mathcal{D}}:=\{(V^{(i)},U^{(i)})_{i=1}^K\}$ with $K$ samples. The mapping from $V$ to $U$ is denoted as $f:\mathbb{R}^{(N,N,P)} \rightarrow \mathbb{R}^{(N,N,Q)}$. 
In data-driven approaches, ${f}$ is modeled non-informatively as a deep neural network (or other statistical models such as Gaussian Processes).
In physics-based approaches, ${f}$ is modeled by discretizing the governing PDE with numerical solvers such as finite-difference, finite-volume, or finite-element methods. We will briefly review neural networks and finite element analysis in the following two subsections.



\subsection{Neural Networks}

A $L$-layer neural network is a function ${ y(x|W) = f^L(W^L, ... f^2(W^2, f^1(W^1, x))) }$ with parameters $W=\{W^1, W^2 ... W^L\}$ and input $x$. The function $f$ is called activation function, which acts on all components of the input vector. During the training phase, the network weights are determined by minimizing the difference between network output and observations. 
It is found that, given enough nodes, neural networks with non-linear activation function have the potential to approximate any complicated functions~\cite{hornik1991universal}.  However, how to design the network to be more efficient for different problems is always an open question.

\begin{figure} 
  \centering
    \includegraphics[width=0.85\linewidth]{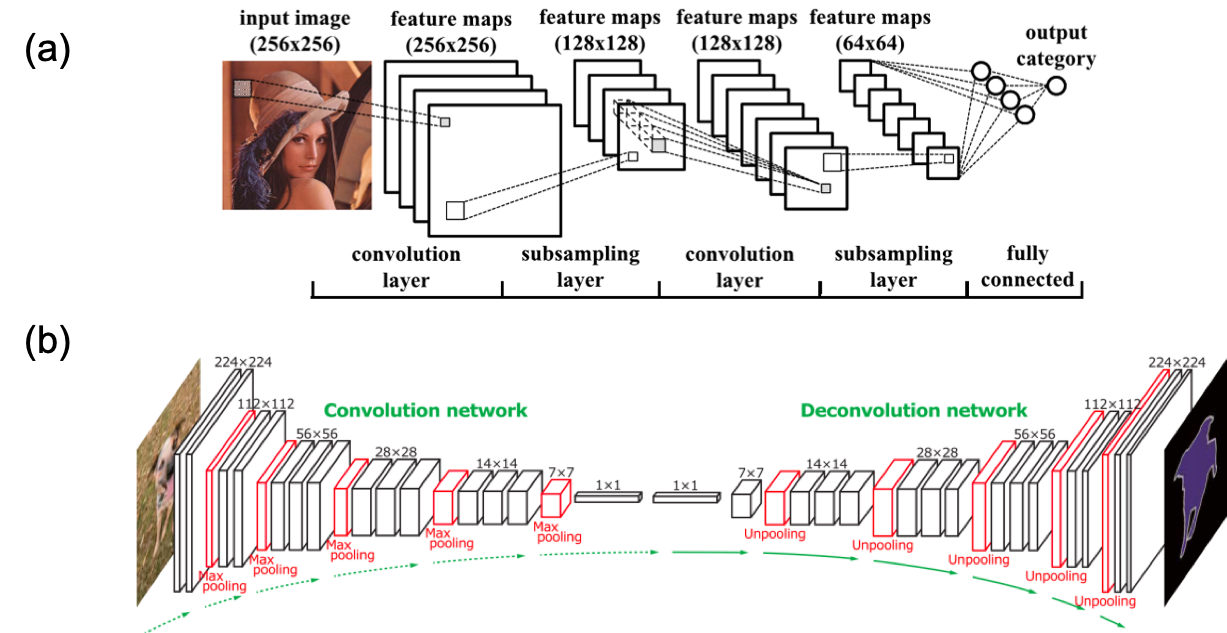}
    \caption{(a): CNN, (b): FCN (figure adopted from \cite{ren2015frcnn})}
\label{fig:CNN}
\end{figure}

A plethora of research has been done to design more efficient and effective neural networks among deep learning and computer vision communities. Some of the biggest breakthroughs can be summarized as:
(1) Replacing some fully connected layers with convolutions \cite{lecun1989lenet} (as shown in Fig.~\ref{fig:CNN}a). In this way, Convolutional Neural Network (CNN) mimics the human visual system and captures the spatial correlations better. It has shown to be very suitable for various vision-based tasks like object recognition \cite{krizhevsky2012imagenet,lecun1989lenet}, detection \cite{girshick2014rcnn,ren2015frcnn}, generation \cite{radford2015dcgan,isola2017image2image}, and segmentation \cite{long2015fcn,ronneberger2015unet}. 
(2) The invention of residual networks (ResNet) \cite{he2016resnet}. Through the short-cut residual connections, Res-Net style network can avoid the notorious ``gradient vanishing'' problem and make the training of network with thousands of layers possible. Compared with previous neural networks, ResNet and its various extensions \cite{huang2017densely,xie2017resnext} can almost always achieve better convergence and higher accuracy. 
{\color{black}(3) The development of one-shot learning algorithms \cite{fei2006oneshot,santoro2016meta}. Based on either Bayesian theory \cite{fei2006oneshot} or external network memory and attention mechanism \cite{santoro2016meta}, these models can be very data efficient and partially mitigates the need for big data for network training.}

Designed to perform semantic segmentation of images, fully Convolutional Network (FCN) is a special type of CNN that only contains convolutional layers \cite{long2015fcn,ronneberger2015unet}. As shown in Fig.~\ref{fig:CNN}b, it takes in images as input and outputs another image of the same resolution with per-pixel label.
Since only convolution operation is involved, FCN is very computationally efficient and can handle inputs of arbitrary size.

\subsection{Finite Element Analysis}
The core idea behind Finite Element Analysis (FEA) is to approximate the potential field with piece-wise lower-order functions~\citep{hughes2012fea}. 
In practice, it involves discretizing the solution domain with smaller meshes, which transforms the original PDE into a system of linear equations:
\begin{equation}
\label{eq:fea_eq}
    K \cdot u = v
\end{equation}
where $v$ and $u$ are the vectors of system loading and response defined on the descritized FEA nodes, and $K$ is the global stiffness matrix which is obtained by assembling all individual element stiffness matrices $K^e$:  
\begin{equation}
\label{eq:element_stiffness}
K^e = \int_{\Delta} B^T C B {\color{black}d\Omega}
\end{equation}
where $C$ is the constitutional matrix depends on the material property, $B$ is the geometry matrix decided by the element shape and order, and $\Delta$ is the finite element. 

While numerous numerical solvers exist for solving Eq.~\ref{eq:fea_eq}, most of them involves iteratively computing the residual~\citep{yang2014SRJ}:
\begin{equation}
\label{eq:itr_sol}
    r = v - K \cdot u
\end{equation}
As an example, the simplest Jacobi solver has the following form:
\begin{equation}
\label{eq:jacobi_itr}
    u_{t+1} = \omega D^{-1} \cdot r_t + u_t
\end{equation}
where $\omega$ is a hyper-parameter, and $D$ is the diagonal part of matrix $K$. 

\subsection{Deep learning with physics}

Recent attempts have been made to predict physics response or parameters with data driven models~\cite{sheikholeslami2019nanofluid,wang2019load_path,finol2018eigen,sosnovik2017neural,li2019predicting_shale}, and several seminal works have been done to build hybrid learning mechanisms with physics knowledge \cite{cang2018cad,hsieh2018ICLR_PDE,li2000gfs_cmame,oishi2017enhance_fea,capuano2019smart_FEA,yu2018physics,lu2017beyond,long2018pde}. 
{\color{black}Early pioneering work has shown that the global or element stiffness matrix can be learnt with neural network for simple systems \cite{li2000gfs_cmame}.}
Based on the optimality condition of topology optimization, efficient optimum topology generators have been designed and trained in \cite{cang2018cad}. 
The performance of FEA is enhanced by utilizing neural networks to learn better integration rule \cite{oishi2017enhance_fea} or element information \cite{capuano2019smart_FEA}.
These works use a neural network as a module under the FEA framework \cite{oishi2017enhance_fea,capuano2019smart_FEA}, which differs from the proposed method which focuses on designing efficient network architectures inspired by FEA. 
Parallel works have been done to improve the convergence and accuracy of finite difference analysis (FDA) solvers to initial value problems (IVP) through learning the optimum filters \cite{hsieh2018ICLR_PDE}, or by building a hybrid model with ODE information hard-coded \cite{yu2018physics}. 
It is found that similarities exist between different FDA solvers and some neural network structures in \cite{lu2017beyond}. Based on this finding, \cite{long2018pde} proposed PDE-Net based on finite difference scheme and reported promising result in system identification. 

Our previous work has shown that the matrix-vector production for FEA can be reduced to a simple convolutional operation for homogeneous material \cite{yao2019fea}. In this paper, this idea is further extended to multi-physics and multi-phase systems.
Our work is similar to PDE-Net~\cite{long2018pde} to some extent, as both networks aim at making use of prior knowledge in PDE and its solvers to build better network architecture. The main difference is that, while PDE-Net is built on FDA for IVP, FEA-Net based on FEA for boundary value problems (BVP). Moreover, we successfully generalized our solver to handle bi-phase materials and can learn a richer material phase information. 
{\color{black} Our work is similar to the FEA with element-by-element technique; however, the proposed FEA convolution operator can be easily learned from data (e.g., analogy of the classical model calibration using inverse FEM analysis). To the best of our knowledge, this is the first time that FEM (with EBE) is expressed into a CNN, which allows the explicit learning of physics (e.g., materials parameters and microstructures) during the network training process. The proposed method bridges FEA and CNNs, which enables the future potential knowledge transfer from a larger deep learning community to the computational mechanics community. }

\section{FEA convolution}
\label{sec:fea_conv}
This section is organized into the following parts: We start by introducing the FEA convolution to model PDE for homogeneous material, and generalize it to handle multi-physics problems in Sec.~\ref{subsec:fea_conv_single_phase}.
Proposed FEA convolution is then further extended to multi-phase problems in Sec.~\ref{subsec:fea_conv_bi_phase}. 
How the gradient of FEA convolution can be obtained is discussed in Sec.~\ref{subsec:gradient}.

\subsection{FEA convolution for multi-physics problem} 
\label{subsec:fea_conv_single_phase}
For physics process, its system loading and response need to satisfy some underlying PDE. Based on finite element analysis, there exists a ''local support property'': The loading at any node is related to only the response at its surrounding nodes. 
Thus, in image space, any pixel value in image $V$ is only related to the pixel values in $U$ at its neighboring region. This relationship is formalized into the following theorem:

\begin{theorem}
The mapping from system response image $U \in R^{(N,N,Q)}$ to system loading image $V \in R^{(N,N,P)}$ can be modeled with a convolution operation for homogeneous material:
\begin{equation}
\label{eq:fea_conv}
    V =  W \circledast U
\end{equation} 
where $\circledast$ and $W \in R^{(R,R,P,Q)}$ denotes the convolution operator and filter, and $R$ can be any odd number larger than 3. 
\end{theorem}

\begin{proof}
We give the proof with single input and output component ($P=Q=1$), which can be extended to other cases naturally. 
Under FEA perspective, the relationship between $U$ and $V$ can be defined by Eq.~\ref{eq:fea_eq}, with element stiffness matrix defined in Eq.~\ref{eq:element_stiffness}. It is worth noting that, if the physics problem is unchanged and the material is homogeneous everywhere in $\Omega$, the constitutional matrix $C$ will be the same for all elements. Furthermore, if the mesh is uniform and of the same order, then the shape matrix $N$ will be the same as well. Under these hypotheses, the element stiffness matrices $K^e$ will be the same everywhere.

\begin{figure} 
  \centering
    \includegraphics[width=0.5\linewidth]{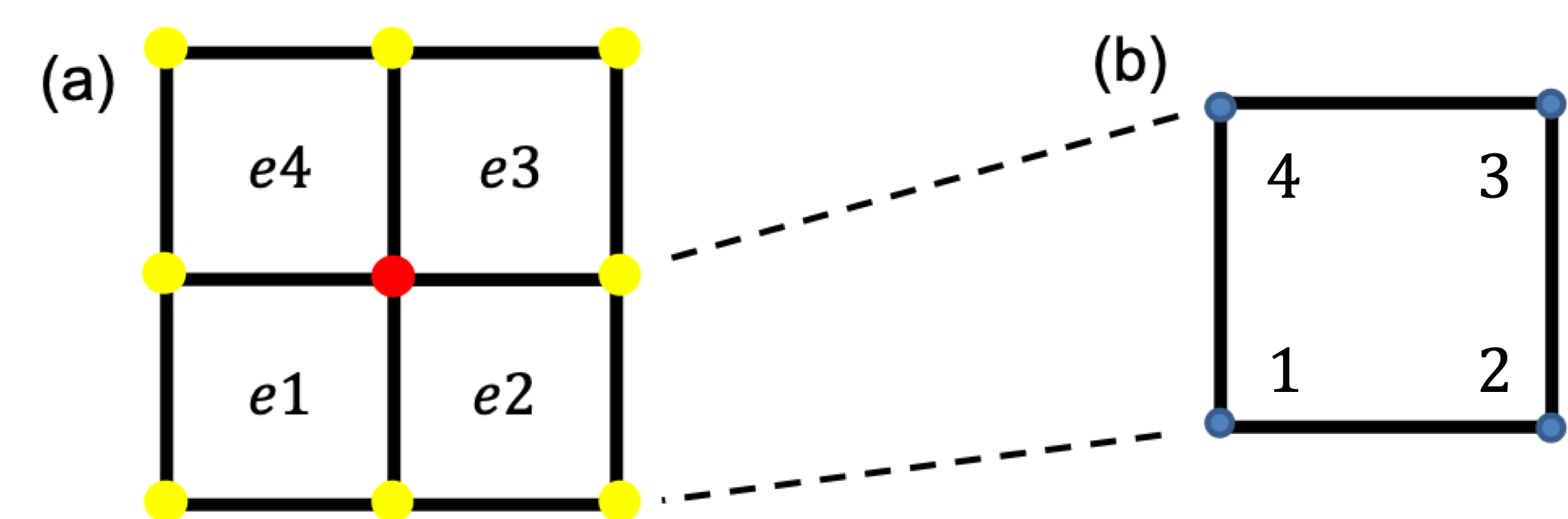}
    \caption{
    (a) The response of $node(i,j)$ (plotted in red) is affected by the loading on the nodes of the four surrounding elements only.
    (b) Node numbering convention for a element.
    }
\label{fig:setup}
\end{figure}

As an example, we use the simplest 4 node linear element to discretize the underlying PDE in this paper. Following the numbering convention in Fig.~\ref{fig:setup}, such discretization will lead us to a system of linear equations:
\begin{equation}
\label{eq:fea_expansion}
\begin{split}
   V_{ij} & = K^{e1}_{13} \cdot U_{i+1,j-1} + K^{e1}_{23} \cdot U_{i+1,j} + K^{e1}_{33} \cdot U_{i,j}  + K^{e1}_{43} \cdot U_{i,j-1} \\
   &+ K^{e2}_{14}\cdot U_{i+1,j} + K^{e2}_{24}\cdot U_{i+1,j+1} + K^{e2}_{34}\cdot U_{i,j+1}  + K^{e2}_{44}\cdot U_{i,j} \\
   &+ K^{e3}_{11}\cdot U_{i,j} + K^{e3}_{21}\cdot U_{i,j+1} + K^{e3}_{31}\cdot U_{i-1,j+1}  + K^{e3}_{41}\cdot U_{i-1,j} \\
   &+ K^{e4}_{12}\cdot U_{i,j-1} + K^{e4}_{22}\cdot U_{i,j} + K^{e4}_{32}\cdot U_{i-1,j}  + K^{e4}_{42}\cdot U_{i-1,j-1}\\
\end{split}
\end{equation}
where e1 to e4 denotes the four neighbouring elements of $node(i,j)$ as Fig.~\ref{fig:setup}a shows. The subscript of the element stiffness matrix goes from 1 to 4, which corresponds to the node index inside a particular element as Fig.~\ref{fig:setup}b shows.

By comparing Eq.~\ref{eq:fea_conv} and Eq.~\ref{eq:fea_expansion}, we can obtain a 3-by-3 FEA convolution kernel $W$ explicitly:
\begin{equation}
\label{eq:conv_kernel_expression}
  W = 
    \begin{bmatrix}
        K^e_{42}       & K^e_{32}+K^e_{41}       & K^e_{31}\\
        K^e_{43}+K^e_{12}       & K^e_{11}+K^e_{22}+K^e_{33}+K^e_{44}       & K^e_{34}+K^e_{21}\\
        K^e_{13}       & K^e_{23}+K^e_{14}       & K^e_{24}
    \end{bmatrix}
\end{equation} 

Moreover, the filter kernels for higher-order elements can also be obtained based on its element stiffness matrix. For example, a 5-by-5 kernel can be obtained with second-order elements, and a 7-by-7 kernel can be obtained with third-order elements.
\end{proof}

In the rest of this paper, we assume that the filters have a spatial size of 3 by 3 for simplicity.
All terms in $W$ all have their physics meaning. For example, $W_{22}$ (and $W_{11}$) represents the loading at a particular pixel, when there is only a unit response at that pixel itself exist (or the upper left of that pixel). The convolution kernel can be obtained in a closed-form if the physics is perfectly known, otherwise we will need to learn it from data. 
In Sec.~\ref{sec:experiment}, we will verify the proposed kernel and learning approach by comparing the learned filter with its analytical value. 
Below we give examples of the analytical FEA convolution kernel for some known physics problems.

\begin{exmp}
\label{example:thermal}
The analytical FEA convolution kernel for thermal conduction problems is:
\begin{equation}
\label{eq:w_therm}
\begin{split}
    W^{tt} &= \frac{\kappa}{3}
    \begin{bmatrix}
    1  &1  &1\\
    1 &-8 &1\\
    1 &1 &1
    \end{bmatrix}\\
\end{split}
\end{equation}
where $\kappa$ is the thermal conductivity coefficient. 
\end{exmp}
The derivation of this kernel is in \ref{appendix:thermal_kernel}. It is interesting to note that $W^{tt}$ is actually a Laplacian filter. The reason for this is that the governing equation for thermal conduction is Laplacian (Poisson) equation.

\begin{exmp}
The convolution for elasticity has two input and output channels, representing the x- and y- directional components of loading and response respectively. Thus there are four different filters in total.
The filters correspond to the interactions from the same input and output channels are:
\begin{subequations}
\begin{equation}
\label{eq:w_elast}
\begin{split}
    \big(W^{xx}\big)^T = W^{yy} &= \frac{E}{4(1-\nu^2)}
    \begin{bmatrix}
        -(1-\nu/3)        & -2(1+\nu/3)       & -(1-\nu/3)\\
        4\nu/3           & 8(1-\nu/3)        & 4\nu/3\\
        -(1-\nu/3)        & -2(1+\nu/3)       & -(1-\nu/3)
    \end{bmatrix}\\
\end{split}
\end{equation}
where $E$ and $\nu$ are Young's modulus an Poisson's ratio respectively.
And the coupling terms between the two channels are:
 \begin{equation}
\label{eq:w_elast_couple}
\begin{split}
    W^{xy} = W^{yx} &= \frac{E}{2(1-\nu)}
    \begin{bmatrix}
        1       & 0       & -1\\
        0       & 0       & 0\\
        -1       & 0       & 1
    \end{bmatrix}\\
\end{split}
\end{equation}
\end{subequations}
\end{exmp}
The derivation of these kernels is included in \ref{appendix:elasticity_kernel}. Again, FEA convolutional kernel for elasticity also exhibits many interesting properties: The non-coupling terms are symmetric along both axis, and they are just rotated versions of each other. The coupling filters is nonzero only at diagonals, because axial loading does not cause any shear effects for homogeneous material. 

\begin{exmp}
\label{example:couple}
The analytical FEA convolution kernel for the coupling effect between thermal and elasticity is:
\begin{equation}
\begin{split}
    -(W^{xt})^T = W^{yt} &= \frac{\alpha E}{6(1 - \nu)}
    \begin{bmatrix}
        1       & 4       & 1\\
        0       & 0       & 0\\
        -1       & -4       & -1
    \end{bmatrix}\\
    W^{tx} = W^{ty} &= 0
\end{split}
\end{equation}
where $\alpha$ is the thermal expansion coefficient.
\end{exmp}

Since the coupling between thermal and elasticity is a one-way coupling, $W^{tx}$ and $W^{ty}$ are all zeros. The detailed derivation of the thermoelasticity coupling kernel is included in \ref{appendix:coupling}.

\begin{exmp}
The governing equation to thermoelasticity problem can be expressed into a convolution form of:
\begin{equation}
\label{eq:coupling}
\begin{split}
    \begin{bmatrix}
        V^{x}\\
        V^{y}\\
        V^{t}\\
    \end{bmatrix} = 
    \begin{bmatrix}
        W^{xx}       & W^{xy}       & W^{xt}\\
        W^{yx}       & W^{yy}       & W^{yt}\\
        W^{tx}       & W^{ty}       & W^{tt}\\
    \end{bmatrix} 
    \circledast 
    \begin{bmatrix}
        U^{x}\\
        U^{y}\\
        U^{t}\\
    \end{bmatrix} 
\end{split}
\end{equation}
\end{exmp}

This can be obtained by simply combining Example \ref{example:thermal} to  Example \ref{example:couple} together. Both input and output image has three channels, which corresponds to x- and y-directional mechanical components and thermal component. Thus, the overall filter for the coupling field is a tensor with dimension $R^{(3,3,3,3)}$.

\subsection{FEA convolution for multi-phase problem}
\label{subsec:fea_conv_bi_phase}
Now we demonstrate how this idea can be extended to multi-phase problems. Without loss of generality, we use bi-phase elasticity as an example. A binary-valued image $H$ is first introduced to represent the material phase. The pixel value of $H$ represents which material phase exists at the specific spatial location (element). The resolution of $H$ is set to $N-1$ by $N-1$, as the number of elements is one less than the number of nodes in FEA with linear element.  

For bi-phase elasticity, the loading image $V \in R^{(N,N,2)}$ would be related to both the response image $U$ and the phase image $H$. We define FEA convolution for bi-phase material as:
\begin{equation}
\label{eq:mask_conv_short}
V=\Theta (\rho) \otimes (U,H)
\end{equation}
where $\otimes$ and $\Theta \in R^{(2,P,Q,S,S)}$ are the FEA convolution operator and kernel for bi-phase elasticity. For 2D elasticity, we have  $P=Q=2$, which represents the x- and y- component. And for linear element we have $S=4$, which represents there are four nodes in each element. And $\Theta$ is further assumed to be depended on some physics hyper-parameters, e.g. $\rho = \{E, \nu\}$ for elasticity.

Following the numbering convention in Fig.~\ref{fig:setup}, with four-node linear finite element of the same size, the relationship between phase, response, and loading images can be obtained with FEA as:
\begin{equation}
\label{eq:mask_conv}
\begin{split}
    V_{ij}^{q} = \sum_{h=0}^1 \Big(& C_{hijp}^{e1} \cdot (h + (-1)^h H_{i,j-1})
           + C_{hijp}^{e2} \cdot (h + (-1)^h H_{i,j})\\
           +& C_{hijp}^{e3} \cdot (h + (-1)^h H_{i-1,j})
           + C_{hijp}^{e4}\cdot (h + (-1)^h  H_{i-1,j-1}) \Big)
\end{split}
\end{equation}
where $e1$ to $e4$ still denotes the four neighbouring elements of $node(i,j)$ as Fig.~\ref{fig:setup}a shows. $h$ is either 0 or 1, representing which material phase is under consideration. And $C_{hijp}$ is obtained from $U$ and $W$ as:
\begin{subequations}
\label{eq:mask_conv_terms}
\begin{equation}
\begin{aligned}
    C_{hijp}^{e1} = \sum_{q=1}^{Q} &\Big(\Theta^{hpq}_{31}\cdot U_{i+1,j-1}^{q} + \Theta^{hpq}_{32}\cdot U_{i+1,j}^{q} + \Theta^{hpq}_{33}\cdot U_{i,j}^{q}  + \Theta^{hpq}_{34}\cdot U_{i,j-1}^{q} \Big)\\
\end{aligned}
\end{equation}
\begin{equation}
\begin{aligned}
    C_{hijp}^{e2} = \sum_{q=1}^{Q} &\Big(\Theta^{hpq}_{41}\cdot U_{i+1,j}^{q} + \Theta^{hpq}_{42}\cdot U_{i+1,j+1}^{q} + \Theta^{hpq}_{43}\cdot U_{i,j+1}^{q}  + \Theta^{hpq}_{44}\cdot U_{i,j}^{q}  \Big)\\
\end{aligned}
\end{equation}
\begin{equation}
\begin{aligned}
    C_{hijp}^{e3} = \sum_{q=1}^{Q} &\Big(\Theta^{hpq}_{11}\cdot U_{i,j}^{q} + \Theta^{hpq}_{12}\cdot U_{i,j+1}^{q}  + \Theta^{hpq}_{13}\cdot U_{i-1,j+1}^{q}  + \Theta^{hpq}_{14}\cdot U_{i-1,j}^{q}  \Big)\\
\end{aligned}
\end{equation}
\begin{equation}
\begin{aligned}
    C_{hijp}^{e4} = \sum_{q=1}^{Q} &\Big(\Theta^{hpq}_{21}\cdot U_{i,j-1}^{q} + \Theta^{hpq}_{22}\cdot U_{i,j}^{q}  + \Theta^{hpq}_{23}\cdot U_{i-1,j}^{q}  + \Theta^{hpq}_{24}\cdot U_{i-1,j-1}^{q}  \Big)\\
\end{aligned}
\end{equation}
\end{subequations}

\begin{figure} 
  \centering
    \includegraphics[width=0.8\linewidth]{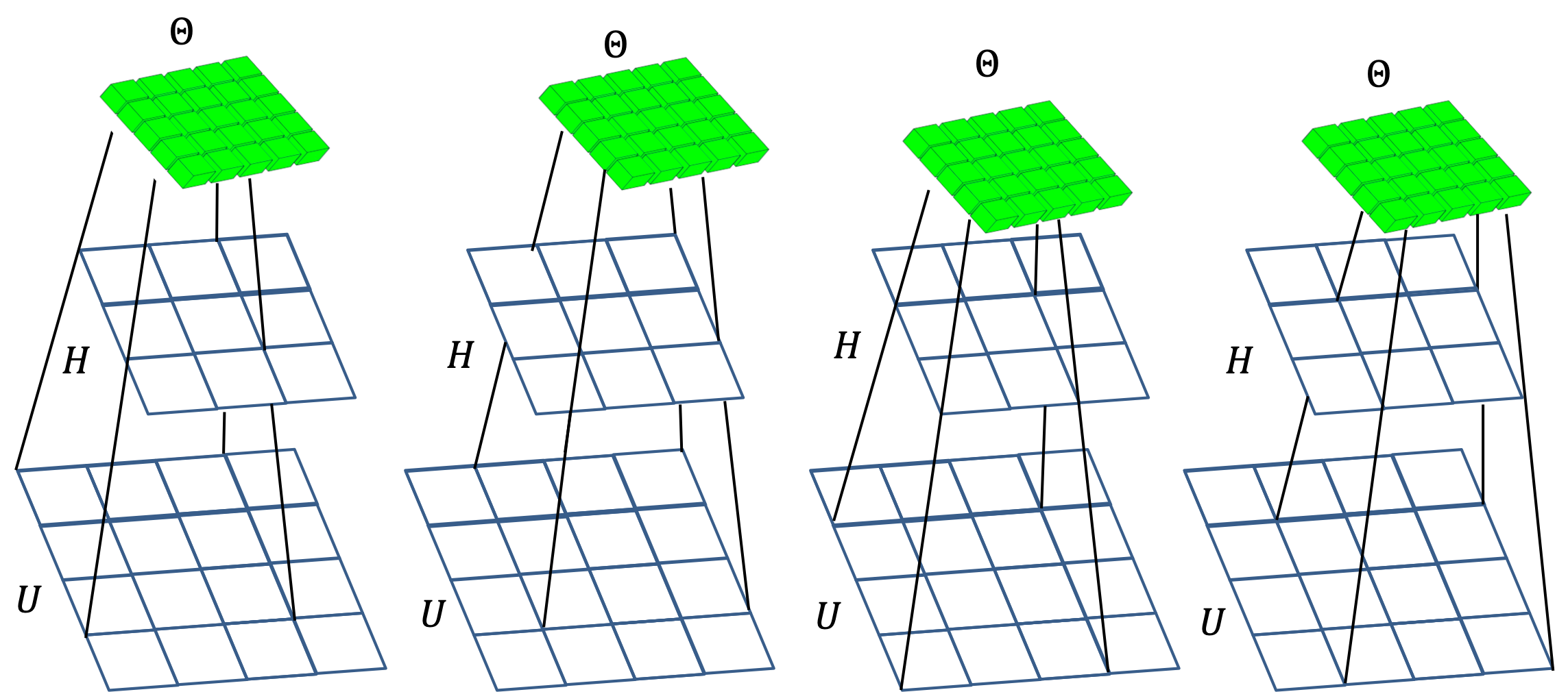}
    \caption{Illustration of the Bi-phase FEA Convolution. Both $H$ and $U$ is involved in FEA convolution. While $\Theta$ is fixed, $H$ and $U$ involved in the computation will be shifted by 1 pixel each time.  
    }
\label{fig:mask_conv}
\end{figure}

The FEA convolution for bi-phase material is illustrated in Fig.~\ref{fig:mask_conv}. The filter $\Theta$ is applied to both a 2-by-2 region in $H$ and a 3-by-3 region in $U$ at the same location. Similar to conventional convolution, the filter will be shifted by one pixel every time and applied to different regions of the images. 
For homogeneous material, Eq.~\ref{eq:mask_conv} will be reduced to Eq.~\ref{eq:fea_conv} by setting $H \equiv 1$.
 
\begin{exmp}
The bi-phase FEA convolutional kernel $\Theta$ for elasticity can be obtained by splitting the element stiffness matrix (which can be found in Eq.~\ref{eq:elast_stiffness} in the Appendix):
\begin{subequations}
\label{eq:w_elast_xx}
\begin{equation}
\begin{split}
\Theta^{h00} &= 
    \frac{E_h}{12(1-\nu_h^2)}
    \begin{bmatrix}
    -2\nu_h+6 & -\nu_h-1 & \nu_h-1 & 2\nu_h \\
    -\nu_h-1 & -2\nu_h+6 & 2\nu_h &\nu_h -1 \\
    \nu_h-1 & 2\nu_h & -2\nu_h+6 & -\nu_h-1\\
    2\nu_h &\nu_h -1 & -\nu_h-1 & -2\nu_h+6\\
    \end{bmatrix}\\
\end{split}
\end{equation}
\begin{equation}
\begin{split}
\Theta^{h11} &= 
    \frac{E_h}{16(1-\nu_h^2)}
    \begin{bmatrix}
    -2\nu_h+6 & 2\nu_h & \nu_h-1 &- \nu_h-1 \\
    2\nu_h & -2\nu_h+6 & -\nu_h-1 &\nu_h -1 \\
    \nu_h-1 & -\nu_h-1 & -2\nu_h+6 & 2\nu_h \\
    -\nu_h-1 &\nu_h -1 & 2\nu_h & -2\nu_h+6\\
    \end{bmatrix}\\
\end{split}
\end{equation}
\end{subequations}
where $E_h$ and $\nu_h$ represents the Young's Modulus and Poisson ration for different material phases. And the coupling filters between different input and output channels are:
\begin{subequations}
\label{eq:w_elast_xy}
\begin{equation}
\begin{split}
\Theta^{h01} &= 
    \frac{E_h}{8(1-\nu_h^2)}
    \begin{bmatrix}
    \nu_h+1 & 1-3\nu_h & -\nu_h-1 & 3\nu_h-1 \\
    3\nu_h-1 & -\nu_h-1 & 1-3\nu_h & \nu_h+1\\
    -\nu_h-1 & 3\nu_h-1 & \nu_h+1 & 1-3\nu_h\\
    1-3\nu_h & \nu_h+1 & 3\nu_h-1 & -\nu_h-1\\
    \end{bmatrix}\\
\end{split}
\end{equation}
\begin{equation}
\begin{split}
\Theta^{h01} &= 
    \frac{E_h}{8(1-\nu_h^2)}
    \begin{bmatrix}
    \nu_h+1 & 3\nu_h-1 & -\nu_h-1 & 1-3\nu_h \\
    1-3\nu_h & -\nu_h-1 & 3\nu_h-1 & \nu_h+1 \\
    -\nu_h-1 & 1-3\nu_h & \nu_h+1 & 3\nu_h-1 \\
    3\nu_h-1 & \nu_h+1 & 1-3\nu_h & -\nu_h-1 \\
    \end{bmatrix}\\
\end{split}
\end{equation}
\end{subequations}
\end{exmp}

\subsection{Gradient of FEA convolution}
\label{subsec:gradient}
Since FEA convolution for thermoelasticity is actually a standard multi-channel convolution, standard deep learning packages like Tensorflow\footnote{https://www.tensorflow.org/} can be directly used to obtain its gradient and to perform back-propagation~\citep{rumelhart1988back_propogation}.
However, the gradient for bi-phase FEA convolution in Eq.~\ref{eq:mask_conv_short} needs to be explicitly defined for efficient computation.

Since $V$ is a function of $U$, $E$ and $\Theta$ (or its physical parameter $\rho$) in bi-phase convolution, there will be three different partial derivatives w.r.t. $V$ needs to be computed. Based on Eq.~\ref{eq:mask_conv}, the gradient of the output $V$ w.r.t. input $U$ can be derived as:
\begin{equation}
\label{eq:partial_v_partial_u}
\begin{split}
\frac{\partial V}{\partial U_{ij}^{q}} =  \sum_{h=0}^1 \Big(& \sum_{p=1}^{P} \frac{\partial C_{hijp}^{e1}}{\partial U_{ij}^{q}} \cdot (h + (-1)^h H_{i,j-1}) + \sum_{p=1}^{P} \frac{\partial C_{hijp}^{e2}}{\partial U_{ij}^{q}} \cdot (h + (-1)^h H_{i,j})\\
           +& \sum_{p=1}^{P} \frac{\partial C_{hijp}^{e3}}{\partial U_{ij}^{q}} \cdot (h + (-1)^h H_{i-1,j}) + \sum_{p=1}^{P} \frac{\partial C_{hijp}^{e4}}{\partial U_{ij}^{q}} \cdot (h + (-1)^h H_{i-1,j-1}) \Big)
\end{split}
\end{equation}
where:
\begin{subequations}
\begin{equation}
\begin{split}
    \frac{\partial C_{hijp}^{e1}}{\partial U_{ij}^{q}} =&  \Theta^{hpq}_{31} \cdot \hat V_{i+1,j-1}^{p} + \Theta^{hpq}_{32}\cdot \hat V_{i+1,j}^{p} + \Theta^{hpq}_{33} \cdot \hat V_{i,j}^{p}  + \Theta^{hpq}_{34}\cdot \hat V_{i,j-1}^{p}\\
\end{split}
\end{equation}
\begin{equation}
\begin{split}
    \frac{\partial C_{hijp}^{e2}}{\partial U_{ij}^{q}} =& \Theta^{hpq}_{41}\cdot \hat V_{i+1,j}^{p} + \Theta^{hpq}_{42}\cdot \hat V_{i+1,j+1}^{p} + \Theta^{hpq}_{43}\cdot \hat V_{i,j+1}^{p}  + \Theta^{hpq}_{44}\cdot \hat V_{i,j}^{p}\\
\end{split}
\end{equation}
\begin{equation}
\begin{split}
    \frac{\partial C_{hijp}^{e3}}{\partial U_{ij}^{q}} =&  \Theta^{hpq}_{11}\cdot \hat V_{i,j}^{p} + \Theta^{hpq}_{12}\cdot \hat V_{i,j+1}^{p} + \Theta^{hpq}_{13}\cdot \hat V_{i-1,j+1}^{p}  + \Theta^{hpq}_{14}\cdot \hat V_{i-1,j}^{p} \\
\end{split}
\end{equation}
\begin{equation}
\begin{split}
    \frac{\partial C_{hijp}^{e4}}{\partial U_{ij}^{q}} =& \Theta^{hpq}_{21}\cdot \hat V_{i,j-1}^{p} + \Theta^{hpq}_{22}\cdot \hat V_{i,j}^{p} + \Theta^{hpq}_{23}\cdot \hat V_{i-1,j}^{p}  + \Theta^{hpq}_{24}\cdot \hat V_{i-1,j-1}^{p}\\
\end{split}
\end{equation}
\end{subequations}
And $\hat V$ is the gradient propagated to this FEA convolution during backpropagation, if there are several FEA convolutions stacked upon each other. The value of $\hat V$ is set to 1 if there is no external gradient passing in.

The gradient of the output $V$ w.r.t. to the material phase $H$ in Eq.~\ref{eq:mask_conv} is relatively simpler:
\begin{equation}
\label{eq:df_dx}
\begin{split}
\frac{\partial V}{\partial H_{ij}} =  \sum_{h=0}^1  \sum_{p=1}^{P}& \Big(C_{hijp}^{e1} (h-(-1)^h)\cdot \hat V_{i,j-1}^{p} + C_{hijp}^{e2} (h-(-1)^h)\cdot \hat V_{i,j}^{p} \\
+ &C_{hijp}^{e3} (h-(-1)^h)\cdot \hat V_{i-1,j}^{p} + C_{hijp}^{e4}(h-(-1)^h)\cdot \hat V_{i-1,j-1}^{p} \Big)
\end{split}
\end{equation} 

Recall that we have assumed that the physics interaction $W_{}$ is related to a set of hidden physics parameters $\rho \in R^Q$, then the gradient of the output $F$ in Eq.~\ref{eq:mask_conv} w.r.t. to $\rho$ can be obtained as:
\begin{equation}
\label{eq:partial_v_partial_rho}
\begin{split}
\frac{\partial V}{\partial \rho} =  \sum_{h=0}^1 \sum_{i=1}^{N} \sum_{j=1}^{N} & \sum_{p=1}^{P}
\Big(
 \frac{\partial C_{hijp}^{e1}}{\partial \rho} \cdot (h + (-1)^h H_{i,j-1}) 
+ \frac{\partial C_{hijp}^{e2}}{\partial \rho} \cdot (h + (-1)^h H_{i,j})\\
+& \frac{\partial C_{hijp}^{e3}}{\partial \rho} \cdot (h + (-1)^h H_{i-1,j}) 
+ \frac{\partial C_{hijp}^{e4}}{\partial \rho} \cdot (h + (-1)^h H_{i-1,j-1})
\Big)
\end{split}
\end{equation} 
where:
\begin{subequations}
\label{eq:partial_w_partial_rho_terms}
\begin{equation}
\begin{split}
    \frac{\partial C_{hijp}^{e1}}{\partial \rho} = \sum_{p=1}^{P} \Big(&
    \frac{\partial \Theta^{hpq}_{31}}{\partial \rho}\cdot \hat V_{i+1,j-1}^{p} 
    + \frac{\partial \Theta^{hpq}_{32}}{\partial \rho}\cdot \hat V_{i+1,j}^{p} 
    + \frac{\partial \Theta^{hpq}_{33}}{\partial \rho}\cdot \hat V_{i,j}^{p}  
    + \frac{\partial \Theta^{hpq}_{34}}{\partial \rho}\cdot \hat V_{i,j-1}^{p} \Big)\\
\end{split}
\end{equation}
\begin{equation}
\begin{split}
    \frac{\partial C_{hijp}^{e2}}{\partial \rho} = \sum_{p=1}^{P} \Big(&
    \frac{\partial \Theta^{hpq}_{41}}{\partial \rho} \cdot \hat V_{i+1,j}^{p}
    + \frac{\partial \Theta^{hpq}_{42}}{\partial \rho} \cdot \hat V_{i+1,j+1}^{p} 
    + \frac{\partial \Theta^{hpq}_{43} }{\partial \rho} \cdot \hat V_{i,j+1}^{p} 
    + \frac{\partial \Theta^{hpq}_{44}}{\partial \rho} \cdot \hat V_{i,j}^{p}  \Big)\\
\end{split}
\end{equation}
\begin{equation}
\begin{split}
    \frac{\partial C_{hijp}^{e3}}{\partial \rho} = \sum_{p=1}^{P} \Big(&
    \frac{\partial \Theta^{hpq}_{11}}{\partial \rho} \cdot \hat V_{i,j}^{p} 
    + \frac{\partial \Theta^{hpq}_{12}}{\partial \rho} \cdot \hat V_{i,j+1}^{p} 
    + \frac{\partial \Theta^{hpq}_{13}}{\partial \rho} \cdot \hat V_{i-1,j+1}^{p}  
    + \frac{\partial \Theta^{hpq}_{14}}{\partial \rho} \cdot \hat V_{i-1,j}^{p}  \Big)\\
\end{split}
\end{equation}
\begin{equation}
\begin{split}
    \frac{\partial C_{hijp}^{e4}}{\partial \rho} = \sum_{p=1}^{P} \Big(&
    \frac{\partial \Theta^{hpq}_{21}}{\partial \rho} \cdot \hat V_{i,j-1}^{p} 
    + \frac{\partial \Theta^{hpq}_{22}}{\partial \rho} \cdot \hat V_{i,j}^{p} 
    + \frac{\partial \Theta^{hpq}_{23}}{\partial \rho} \cdot \hat V_{i-1,j}^{p}  
    + \frac{\partial \Theta^{hpq}_{24}}{\partial \rho} \cdot \hat V_{i-1,j-1}^{p}  \Big)\\
\end{split}
\end{equation}
\end{subequations}
If nothing is known about the underlying physics, we can just set $\rho$ to $\Theta$ itself.
If we have some prior knowledge on the underlying physics, for example we know it is an elasticity problem, then the computation of $\partial \Theta / \partial \rho$ can be obtained from Eq.~\ref{eq:w_elast_xx} and Eq.~\ref{eq:w_elast_xy}. 
Furthermore, if we know the material is homogeneous, Eq.~\ref{eq:partial_v_partial_u} to \ref{eq:partial_v_partial_rho} can be largely simplified into the gradient of conventional 2D convolutions. 

\section{FEA-Net}
\label{sec:fea_net}
This section is divided into two parts, where we build FEA-Net for multi-physics and multi-phase problems respectively. 
To maximize the efficiency, different network architectures are designed for learning and inference: We model the inverse mapping from system response to its corresponding loading during the training stage, and another network architecture is built to map the system loading to response during the inference stage.

\subsection{FEA-Net for multi-physics problems}
\label{sec:multi_physics_learning}

\begin{figure} 
  \centering
    \includegraphics[width=0.78\linewidth]{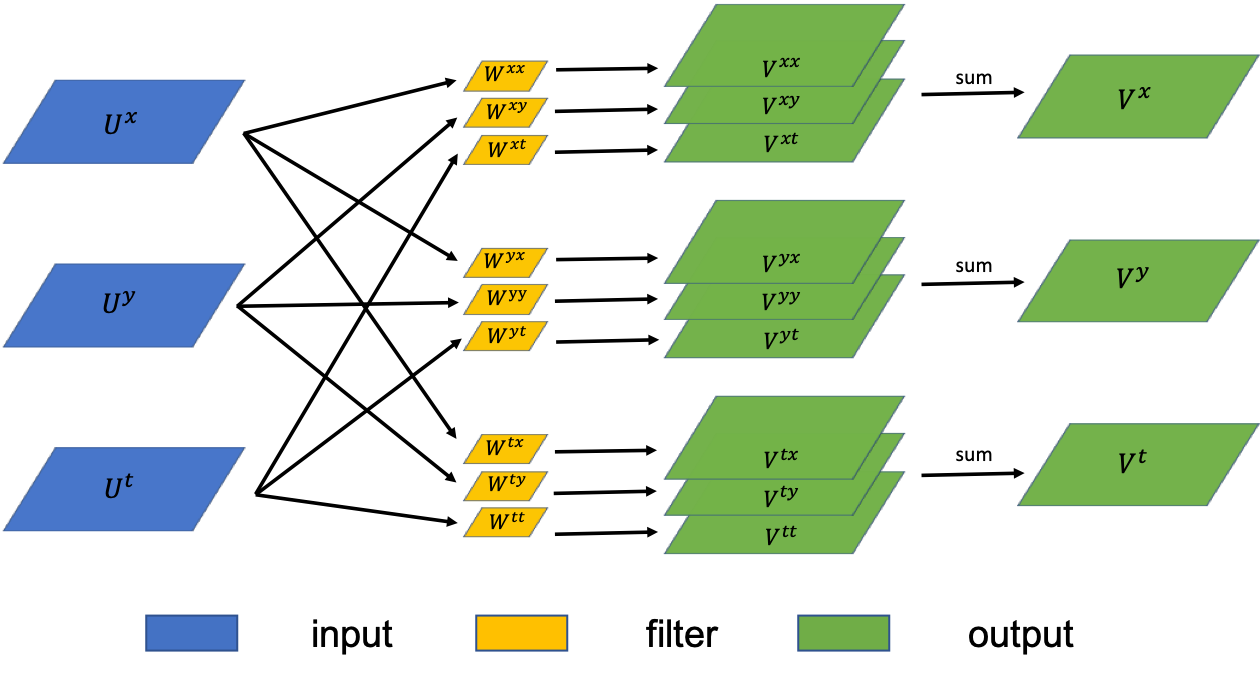}
    \caption{Learning Architecture for Homogeneous Thermoelasticity.}
\label{fig:g_net}
\end{figure}

We use homogeneous thermoelasticity as an example to demonstrate how to design the learning and inference architecture for multi-physics based on the FEA convolution. 
For thermoelasticity, $V \in R^{(N,N,3)}$ (and $U \in R^{(N,N,3)}$) has three channels: x- and y- directional mechanical loading (response) and heat flux (temperature). 
As defined in Eq.~\ref{eq:coupling}, the mapping from $U$ to $V$ is a convolutional operation:
\begin{equation}
V=W\circledast U := h(U,W)
\end{equation}
with $W \in R^{(3,3,3,3)}$. This relationship can be further expressed into a single-layer network with linear activation as illustrated in Fig.~\ref{fig:g_net}. The input and output to the network are the response image $U$ and the predicted loading image $V$ respectively.

Given a training dataset $\mathcal{D}$, the optimum filter $W$ can be obtained by minimizing the difference between the observed system loading $V$ and the predicted loading:
\begin{equation}
\label{eq:qudratic}
\begin{split}
&W^* =  \argmin_{W} \mathbb{E}_{(V,U) \sim \mathcal{D}} L\Big( V, h(U,W) \Big)
\end{split}
\end{equation}
where $L(\cdot, \cdot): \mathbb{R}^{(N, N, 3)} \times \mathbb{R}^{(N, N, 3)} \rightarrow \mathbb{R}$ is a pre-defined loss function, which is chosen as the $L_2$ norm in this paper. FEA convolution will extract the information of the governing PDE during training process.

Once FEA convolution has been trained, we can use it to construct the mapping from $V$ to $U$ and predict the system response when a new loading is applied. 
The core idea is to transform the iterative solvers (as in Eq.~\ref{eq:itr_sol}) into a convolutional network based on the FEA convolution. We will demonstrate with the basic Jacobi solver (as in Eq.~\ref{eq:jacobi_itr}) for its simplicity; however, it is worth noting that the proposed method can be applied with other more advanced iterative solvers as well.

Physically, the diagonal matrix $D$ in Eq.~\ref{eq:jacobi_itr} corresponds to the interaction between $U^q_{ij}$ ($q$-th response component at $node(i,j)$) and $V^q_{ij}$ ($q$-th loading component at $node(i,j)$). This interaction can actually be expressed with $W^{xx}_{22}$, $W^{yy}_{22}$, or $W^{tt}_{22}$ for thermoelasticity.  
Thus, the matrix-vector production $D^{-1} \cdot x$ can be reformulated into an element-wise production $P*X$, with $*$ denotes the element-wise operator.  

\begin{figure} 
  \centering
    \includegraphics[width=0.75\linewidth]{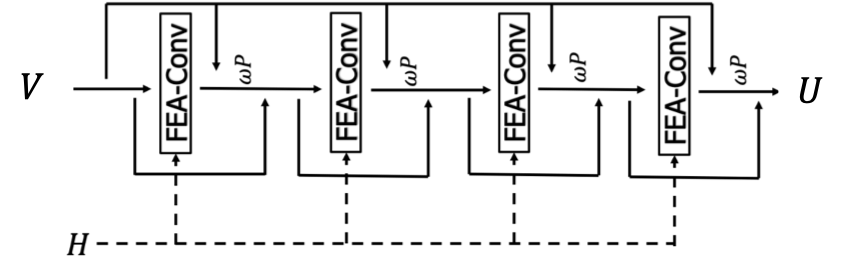}
    \caption{Illustration of FEA-Net. Dash connection is for bi-phase problem only.}
\label{fig:jacobi_net}
\end{figure}

We further define boundary condition operator $\mathcal{B}$, which specifies the Dirichlet boundary condition on $\Gamma$. Operator $\mathcal{B}$ will reset the value $u$ on $\Gamma$ to ground-truth. 
By substituting Eq.~\ref{eq:fea_conv} into Eq.~\ref{eq:jacobi_itr} and apply the boundary condition operator, we have:
\begin{equation}
\label{eq:jacobi_conv_multiphysics}
U_{t+1} = \mathcal{B}\Big( \omega *  P * (V - W \circledast U_{t}) + U^t \Big) 
\end{equation}
The derivation detail can be found in Appendix~\ref{appendix:inference_convergence}. Because most of the computation of Eq.~\ref{eq:jacobi_conv_multiphysics} lies in computing the FEA convolution, it can be viewed as stacking FEA convolutions upon each other. By setting the initial guess $U_0 = V$, Eq.~\ref{eq:jacobi_conv_multiphysics} can be further visualized as a convolutional neural network (as in Fig.~\ref{fig:jacobi_net}). 

The resultant network architecture is similar to both Fully Convolutional Network (FCN) \citep{long2015fcn} and the cutting-edge densely connected ResNet \citep{huang2017densely}, as it is composed of only convolutions and has "short-cuts" across different layers. Similar to FCN, since no fully connected layer is involved, FEA-Net can handle inputs of different size without any problem. 
Most importantly, aside from the similarity on the surface, physics knowledge is inherently embedded in FEA-Net.
Since FEA-Net is designed based on the fix-point iterative solver, so it has certifiable convergence w.r.t. network depth during inference. 

\begin{pro}
The output of the inference network will converge to the ground-truth with increasing network depth, if the network filters have been learned accurately.
\end{pro}
\begin{proof}
See \ref{appendix:inference_convergence} for proof details. Numerical examples can be found in Sec.~\ref{subsec:verification_inference}.
\end{proof}

\begin{pro}
The network filters can be learned accurately with a single image pair, given no linear correlation between different loading images channels.
\end{pro}
\begin{proof}
See \ref{appendix:data_corelation} for proof details. Numerical examples can be found in Sec.~\ref{subsec:verification_multiphysics}.
\end{proof}

Putting Proposition 1 and Proposition 2 together, it can be seen that our model can perform inference with certifiable convergence with a single training image pair.

\subsection{FEA-Net for multi-phase problems}
\label{sec:multi_phase_learning}
From Sec.~\ref{subsec:fea_conv_bi_phase}, we know that the system loading image $V$, response image $U$, and phase image $H$ should satisfy the following relationship:
\begin{equation}
\label{eq:fea_net_h}
V = \Theta (\rho) \otimes (U, H) := h(U,H,\rho)
\end{equation}
where $\Theta$ is the FEA convolutional filter for bi-phase material, which is parametrized by the physics parameters $\rho$. Depending on the availability of the training data, different learning problems can be formulated:

\begin{problem}
\label{problem:est_phase}
Assume that the material property $\rho$ is known, we wish to learn the material phase image $H$ based on the observed system loading and response pair $(V, U)$.
\end{problem}

{\color{black} This particular situation can happen when the material properties of each phase can be obtained from historical database or from experimental testing, such as indentation testing. The micro-structure information is unknown.}
This training process can be formulated into an optimization problem:
\begin{equation}
\label{eq:est_mask}
\begin{split}
H^* =  \argmin_{H} \mathbb{E}_{(V_i,U_i,\rho_i) \sim \mathcal{D}_1} L\Big(V_i, h(\rho_i,H,U_i)\Big)\\
\end{split}
\end{equation}
where $\mathcal{D}_1$ is the training set that contains the loading and response pair obtained with the same material phase.

\begin{pro}
The material phase $H$ can also be correctly learned in any sub-region $\Phi \subset \Omega$, as long as the material property $\rho$ is known and $(V,U)$ image pair has been observed in $\Phi$.  
\end{pro}

\begin{proof}
See \ref{appendix:bi-phase_phase_estimation} for proof details. Numerical examples can be found in Sec.~\ref{subsec:experiment_est_phase}.
\end{proof}

\begin{problem}
\label{problem:est_rho}
Assume the material phase information $H$ is known, we wish to learn the material properties information $\rho$ based on the observed system loading and response pair $(V, U)$. 
\end{problem}

{\color{black} This particular situation can happen when the material micro-structure information is observed from measurements, such as optical imaging and scanning electron microscope imaging. However, the material properties of each phase are unknown.}
Such training process can be formulated as another optimization problem:
\begin{equation}
\label{eq:est_rho}
\begin{split}
&\rho^* =  \argmin_{\rho} \mathbb{E}_{(V_i,U_i,H_i) \sim \mathcal{D}_2} L\Big( V_i, f (\rho_i,H,U_i)  \Big)
\end{split}
\end{equation}
where $\mathcal{D}_2$ is a different training set, which contains the loading, response, and material phase pair obtained under the same material property. 

\begin{pro}
Given material phase image $H$, only a single image pair $(V,U)$ is needed to estimate the material property correctly on both phases, as long as: (1) $H$ contains both phases, and (2) $V$ has none-zero value
\end{pro}

\begin{proof}
See \ref{appendix:bi-phase_rho_estimation} for proof details. Numerical examples can be found in Sec.~\ref{subsec:experiment_est_property}.
\end{proof}

\begin{problem}
\label{problem:est_joint}
Assume that we know the loading and response $(V, U)$, and we wish to estimate both material property and phase together.
\end{problem}

The joint estimation of both material phase and property can be formulated as:
\begin{equation}
\label{eq:est_mask_rho}
\begin{split}
\rho*, H^* =  \argmin_{\rho, H} \mathbb{E}_{(V_i,U_i) \sim \mathcal{D}_3} L\Big(V_i, f (\rho,H,U_i) \Big)\\
\end{split}
\end{equation}
This is a more difficult problem, and we will empirically show that it is also solvable under our framework.

Furthermore, we can put some constraints on the training process if we know which physics parameters are involved.
As the simplest example, if we roughly know the range of the physics parameter $\rho$, we can perform projected gradient descent by:
\begin{equation}
\label{eq:constrain}
\begin{split}
    \rho &= {clip}(\rho, \rho_l, \rho_u)\\ 
\end{split}
\end{equation}
where $\rho_l$ and $\rho_u$ are the lower and upper bound of $\rho$. If we have a better prior knowledge of the distribution of material property, we can have a tighter constraint to make the training even more efficient.

The inference network structure for multi-phase problems can also be obtained from Eq.\ref{eq:jacobi_itr}. By subtracting the diagonal terms from $\Theta$, the expression of $P$ for bi-phase material can be obtained as:
\begin{equation}
\label{eq:D_mat}
\begin{split}
     {P_{ijn}} = 1 /  \sum_{h=0}^1 \Big(&\Theta^{hpq}_{33} \cdot (h + (-1)^h H_{i,j-1})
           +\Theta^{hpq}_{44} \cdot (h + (-1)^h H_{i,j})\\
           +&\Theta^{hpq}_{11} \cdot (h + (-1)^h H_{i-1,j})
           +\Theta^{hpq}_{22} \cdot (h + (-1)^h  H_{i-1,j-1}) \Big)
\end{split}
\end{equation}
Similar to multi-physics problems, by substituting Eq.~\ref{eq:mask_conv_short} and Eq.~\ref{eq:D_mat} into Eq.~\ref{eq:jacobi_itr} we will have the convolutional form of the Jacobi solver: 
\begin{equation}
\label{eq:jacobi_conv_multiphase}
U_{t+1} = \mathcal{B}\Big( \omega *  P * \big(V - \Theta \otimes (U_{t},H) + U^t \Big)
\end{equation}
As with Eq.~\ref{eq:jacobi_conv_multiphysics}, certifiable convergence w.r.t. network depth can also be obtained with Eq.~\ref{eq:jacobi_conv_multiphase}.

\section{Experiments and results}
\label{sec:experiment}
This section is arranged as follows: 
In the first three parts, we verify our learning algorithm for different problems in multi-physics and multi-phase,  as well as the convergence of our inference architecture. 
In the fourth and fifth part, our method is compared to purely data-based and physics-based model respectively.

\subsection{Verification of learning on multi-physics}
\label{subsec:verification_multiphysics}

The learning of physically meaningful filters for the multi-physics problem is verified in this subsection. We use Eq.~\ref{eq:w_therm} to Eq.~\ref{eq:coupling} to obtain the reference filter value directly from material properties.

In the first experiment, we initial the network with filter values with zeros and train it with a single image pair.
The training data is prepared based on numerical simulation with different material properties: We have Young's modulus $E$ ranging from 0.1 TPa to 0.4 TPa, Poisson ratio ranging from 0.2 to 0.35, thermal conductivity from $10 W/(m\cdot K)$ to $14 W/(m\cdot K)$, and thermal expansion ratio from $11/^{\circ} C$ to $15/^{\circ} C$.
With such setup, the training loss value is approaching zero and the relative $L_2$ prediction error is smaller than $10^{-5}$ for all cases.
Three examples of the learned network filters are shown in Tab.~\ref{tab:w_prediction_rand_mat} with their reference value. It can be seen that the value of the learned filter elements is getting very close to the reference value.

\begin{table} 
\centering
\caption{Examples of the learned filter elements with different training data.}
\begin{tabular}{|C{1.7cm}|C{1.7cm}|C{1.7cm}|C{1.7cm}|C{1cm}|C{2.5cm}|C{2.5cm}|}\hline
  \multicolumn{4}{|c|}{Physics parameters} &  \multirow{2}{*}{Filter} &  \multirow{2}{*}{Reference} &  \multirow{2}{*}{Predicted} \\
  \cline{1-4}
  $E ( TPa)$ & $\nu $& $\kappa ( W/(m\cdot K)) $& $\alpha (10^{-5}/^{\circ} C)$ & & &\\
  \hline
  \multirow{4}{*}{0.23}&  \multirow{4}{*}{0.289}&  \multirow{4}{*}{11.82}&  \multirow{4}{*}{12.92}& 
          $W^{xx}_{11}$ & -56.7054195  & -56.7053258 \\
  & & & & $W^{xy}_{11}$ & 40.4512309  & 40.4511605 \\
  & & & & $W^{tx}_{11}$ & -0.159391382  & -0.159391115 \\
  & & & & $W^{tt}_{11}$ & -4.56652389  & -4.56652390 \\
  \hline
  \multirow{4}{*}{0.196}&  \multirow{4}{*}{0.299}&  \multirow{4}{*}{12.97}&  \multirow{4}{*}{12.96}& 
          $W^{xx}_{11}$ & -48.3777489  & -48.3777004 \\
  & & & & $W^{xy}_{11}$ & 34.8800133  & 34.8799752 \\
  & & & & $W^{tx}_{11}$ & -0.150795392  & -0.150795228 \\
  & & & & $W^{tt}_{11}$ & -5.26968980  & -5.26968981 \\
  \hline
  \multirow{4}{*}{0.228}&  \multirow{4}{*}{0.273}&  \multirow{4}{*}{12.92}&  \multirow{4}{*}{11.82}& 
          $W^{xx}_{11}$ & -55.8988893  & -55.8988283 \\
  & & & & $W^{xy}_{11}$ & 39.1483937  &  39.1483472\\
  & & & & $W^{tx}_{11}$ & -0.168541618  & -0.168541427 \\
  & & & & $W^{tt}_{11}$ & -5.91251479  & -5.91251480 \\
  \hline
\end{tabular}
\label{tab:w_prediction_rand_mat}
\end{table}

In the second experiment, with given training data pair, we fix the material property and vary the filter initialization. The statistics of the learned filter over 100 different random initialization is shown in Tab.~\ref{tab:w_prediction_rand_load}. It can be seen that the reference filter has very nice symmetry property. More importantly, for all the random loading/ response data pair, the learned filter is matching with the reference value very well. 

\begin{table} 
\centering
\caption{Comparison on the reference value of the network filter $W^{xx}$ with predictions on 100 random loading images.}
\begin{tabular}{|C{2cm}|C{3cm}|C{3cm}|C{3cm}|}\hline
\setlength{\tabcolsep}{12pt}
  \multirow{2}{*}{Filter} & \multirow{2}{*}{Reference} & \multicolumn{2}{c|}{Prediction}  \\ \cline{3-4}
  ~ & ~ & mean & std \\
  \hline
  $W^{xx}_{11}$ & -52.2454463 & -52.24507827& 0.00147689  \\
  $W^{xx}_{12}$ & 22.19275595 & 22.19259812 & 0.00062905  \\
  $W^{xx}_{13}$ & -52.2454463 & -52.24509037& 0.00144434\\
  \hline
  $W^{xx}_{21}$ & -126.68364854 & -126.68278956& 0.00347672  \\
  $W^{xx}_{22}$ & 417.96357038  & 417.96069933& 0.01163251 \\
  $W^{xx}_{23}$ & -126.68364854 & -126.68276882&  0.00357171\\
  \hline
  $W^{xx}_{31}$ & -52.2454463  & -52.24508064 & 0.00149684  \\
  $W^{xx}_{32}$ & 22.19275595  & 22.19259831 & 0.0006441 \\
  $W^{xx}_{33}$ & -52.2454463  & -52.24508812 &   0.0014402 \\
  \hline
\end{tabular}
\label{tab:w_prediction_rand_load}
\end{table}

\begin{figure}
    \centering
    \includegraphics[width=0.5\linewidth]{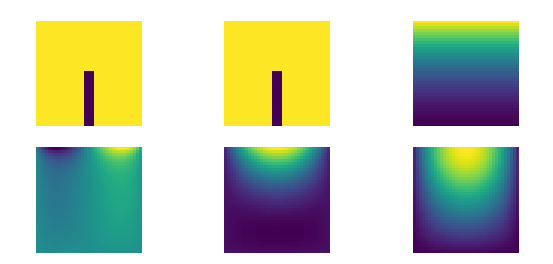}
    \caption{An example of the loading and response image pair that will fail the proposed training. First and second rows are loading and response images. First to last columns are different channels for x- and y-directional mechanical components as well as thermal components.}
    \label{fig:resp_multicollinear}
\end{figure}

The correctness of Proposition 2 has been verified by the previous two experiments. Now we test how the training algorithm performs if the premise of Proposition 2 is violated. 
In this experiment, the loading image/ response data in Fig.~\ref{fig:resp_multicollinear} is used for training, where two loading channels are linearly correlated. This violates the premise of Proposition 2. Although it is difficult to visually tell that the response data has a linear correlation, it is in fact rank deficient (see \ref{appendix:data_corelation} for more details). 
Our network is still able to minimize the loss on such data; however, the learned filter is not unique depending on the initialization. An easy way to mitigate this problem to always apply random loading to obtain the system response. Another alternative solution is to add more prior knowledge to the network, for example, the relationship between network filters and physics parameters (as with solution to Problem1).

{\color{black}
The computational time for one training epoch depends on the implementation and the specific hardware used. As an example, under our Tensorflow implementation on a single Titan XP, it takes 0.06 ms/ 0.10 ms for a single training data with resolution of 50-by-50 to perform a forward/ backward pass.
}

{}

\subsection{Verification of learning on multi-phase}
\label{subsec:verification_multiphase}
In this part, we verify that learning of material phase and property works well under a wide range of different settings like learning rate, initialization, and problem complexity.

We define the relative estimation error of a variable $x$ as:
\begin{equation}
\label{eq:relative_error}
    \epsilon = \frac {|x^{pred} - x^{ref}|_2}  {|x^{ref}|_2}
\end{equation}
The constrain in Eq.~\ref{eq:constrain} is set to $E \in (0, 0.5) TPa$ and $\nu \in (0, 0.5)$. Note this is a very loose constrain that could be satisfied by almost all known material. 

\subsubsection{Material phase estimation}
\label{subsec:experiment_est_phase}

\begin{figure}
  \centering
    \includegraphics[width=0.9\linewidth]{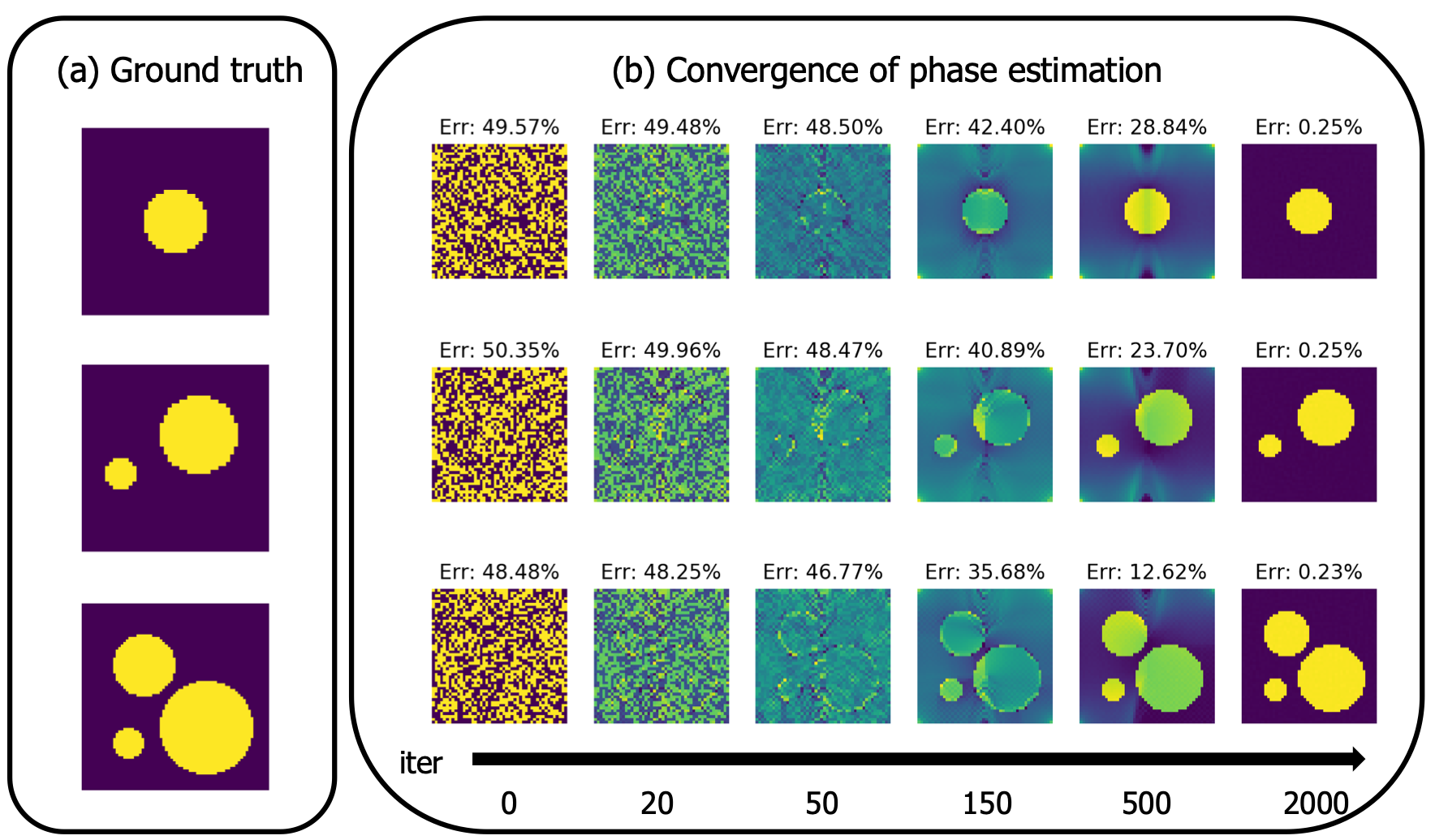}
    \caption{Learning convergence of material phase with different inclusions.
    (a): material phase ground truth
    (b): predicted phase at different iteration starting from random initialization.}
\label{fig:bi-phase_inclusion}
\end{figure}

We first show how our method performs in solving Problem \ref{problem:est_phase} with different complexity of material phase. As shown in Fig.~\ref{fig:bi-phase_inclusion}a, circular inclusions of different size, shape, and location are used in this experiment. Phase images $H$ in this experiment are at a resolution of 50 by 50. 
The inclusion has a Young's modulus of $0.241 TPa$ and a Poisson ratio of $0.36$. The second material has a Young's modulus of $0.2 TPa$ and a Poisson ratio of $0.25$. Still, only a single loading/ response image pair is used for training in this experiment. 

In the first experiment, we show how learning performs on different phase configurations.
The network filter $\Theta$ is initialized with ground truth material property $\rho$, and the material phase image $H$ is initialized randomly. Adam optimizer \citep{kingma2014adam} with a learning rate of $10^{-2}$ is used to run Eq.~\ref{eq:est_mask}. 
The convergence of the material phase is shown in Fig.~\ref{fig:bi-phase_inclusion}b.  The caption shows the relative phase estimation error $\epsilon$, which is dropping below 0.25\% for all phase configurations. 

A second experiment is designed to further evaluate the influence of the image resolution of training data and the learning rate. The single inclusion material shown in Fig.~\ref{fig:bi-phase_inclusion} is used, with image resolution goes from 25 to 50 and 75. The learning rate used is either $10^{-1}$ or $10^{-2}$. We use ``SS'' to abbreviate small image resolution and small learning rate, ``ML'' to abbreviate medium image resolution and large learning rate, and so on. 

\begin{figure}
  \centering
    \includegraphics[width=0.49\linewidth]{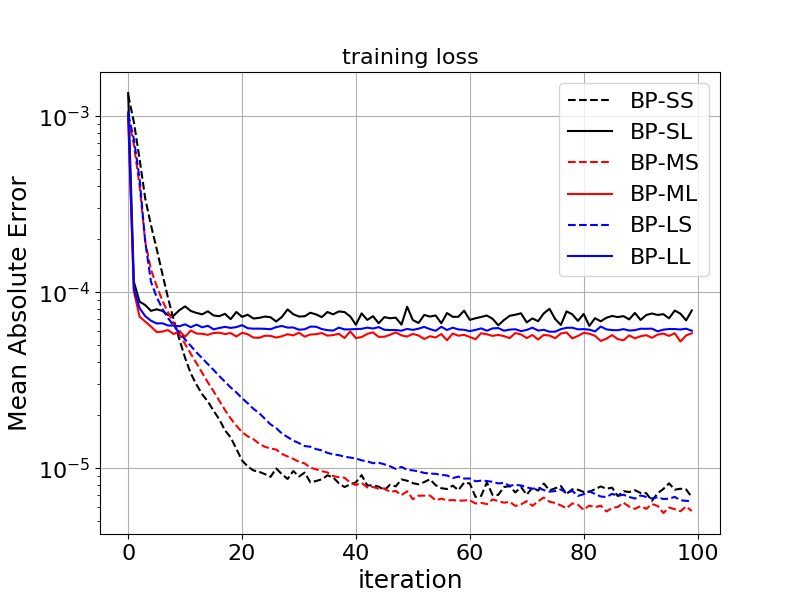}
    \includegraphics[width=0.49\linewidth]{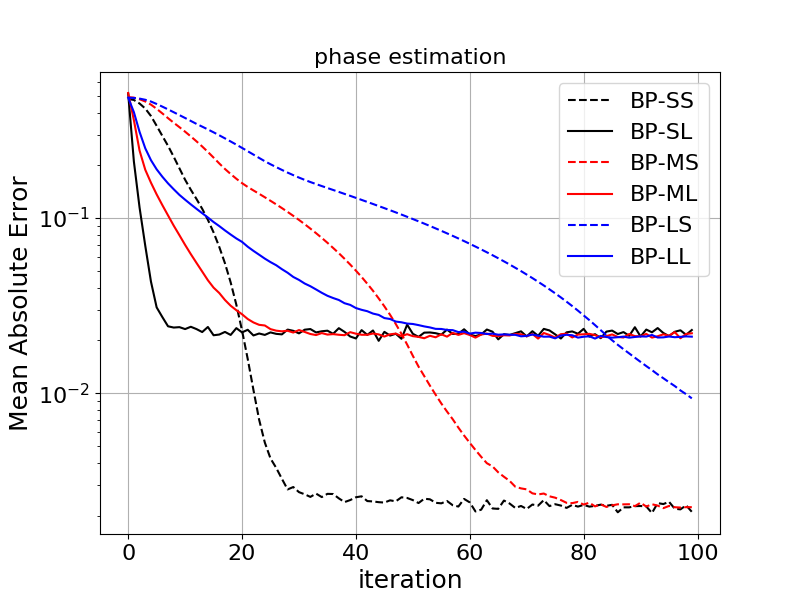}
    \caption{Material Phase Estimation with Different Resolution and Learning rate. (a): Convergence of the loss. (b): Convergence of the prediction error.}
\label{fig:vary_res_and_lr}
\end{figure}

We initialized the material phase image $H$ randomly. The convergence of the network training loss and the prediction error in the material phase is shown in Fig.~\ref{fig:vary_res_and_lr}.
It can be seen that the training is successful under all these settings. And a larger learning rate tends to increase the speed of convergence, with a side effect of a relatively larger error rate. This is because larger learning rate will make it harder to converge to the global optimum. 
Another observation is that, the material phase information is getting much harder to be estimated correctly when the resolution is increased. This is because the search space is getting significantly larger when we increase the image resolution, thus making the optimization problem much more difficult.

\subsubsection{Material property estimation}
\label{subsec:experiment_est_property}

We first test how the learning of material property performs with random initialization. Again, only a single image pair is used for training.
The data is generated with the single circular inclusion phase configuration (as shown by Fig.~\ref{fig:bi-phase_inclusion}a) under a resolution of 50-by-50. The material properties are ${0.241 TPa, 0.36}$ and ${0.2 TPa, 0.25}$ for inclusion and exclusion material. 

\begin{figure}
  \centering
    \includegraphics[width=0.8\linewidth]{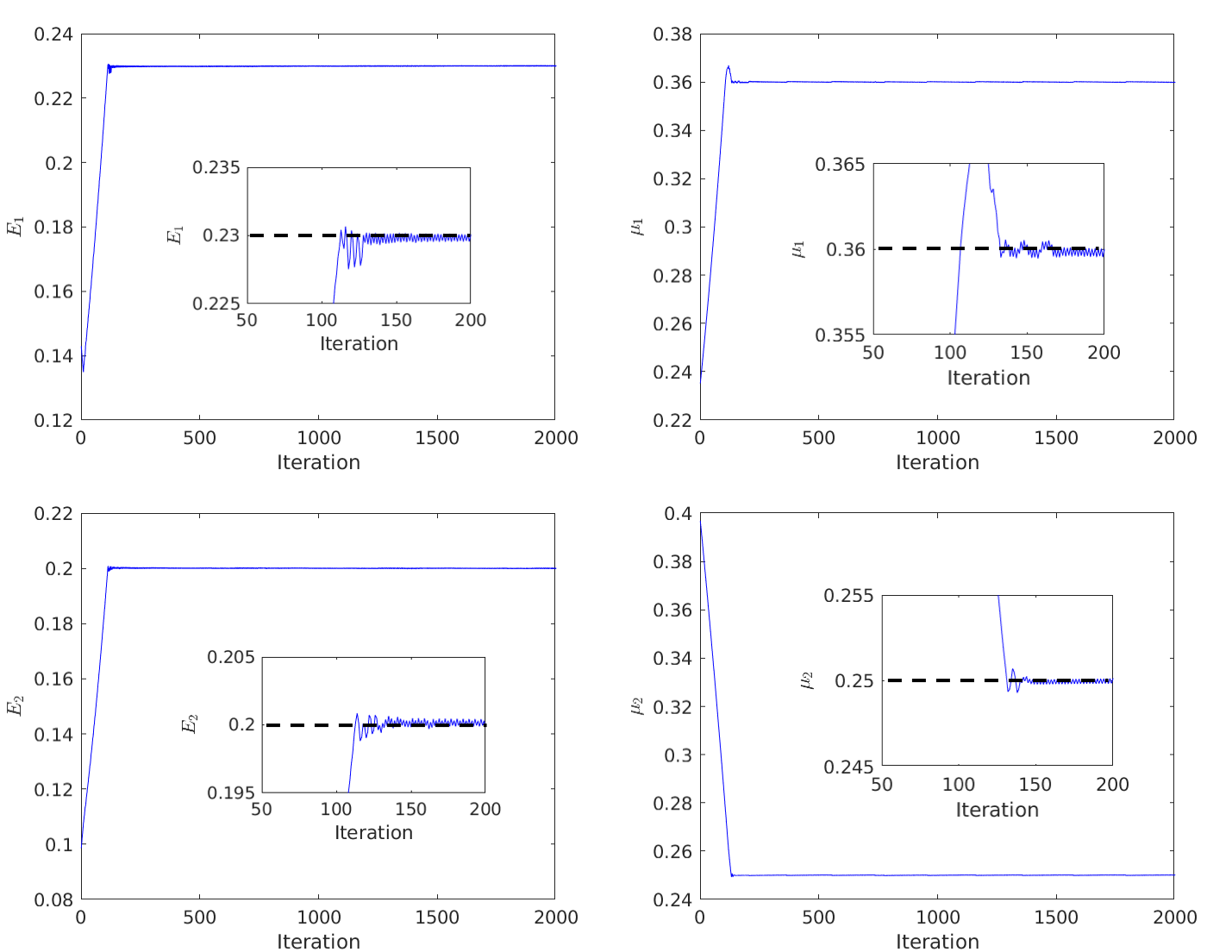}
    \caption{Convergence of Different Material Properties. Reference solution is marked with dashed lines. 
    }
\label{fig:material_property_convergence}
\end{figure}

We initialize a total of 100 different networks, with $E$ ranges from 0 to 0.5 TPa and $\nu$ ranges from 0 to 0.5 for both phases. Adam optimizer with a learning rate of $10^{-3}$ is used to run Eq.~\ref{eq:est_rho}. 
It is worth noting that the training is converging across all 100 samples, which demonstrate that our algorithm is robust towards different initialization. The convergence of one randomly picked network is visualized in Fig.~\ref{fig:material_property_convergence}, where the training process is very stable and all parameters are converging within 150 iterations.

Furthermore, we test the robustness of our training algorithm with data obtained from different material properties. We fix inner material property to be $E=0.2 TPa$ and $\nu=0.25$, and the surrounding material is set to have $E$ vary from 0.05  TPa to 0.441  TPa and $\nu$ vary from 0.1 to 0.4. A total of 25 different data points are generated in this way.
We initialize the network with ground truth material phase, elasticity modulus of $0.01 TPa$ and Poisson ratio of $0.1$ for both phases. Again, Adam optimizer with a learning rate of $10^{-3}$ is used to run Eq.~\ref{eq:est_rho}. The training is converging for all training data to an error rate below 1\%, suggesting that our method is very robust in predicting material properties.



\subsubsection{Joint material phase and property estimation}
\label{subsec:experiment_est_joint}
In this part, we assume both the material phase and property is unknown and will estimate them from the loading/ response pair. 
Material phase images from Fig.~\ref{fig:bi-phase_inclusion}a are used to generate the training data. 
Inclusion (and exterior material) has Young's modulus of 0.241 TPa (and 0.2 TPa) and a Poisson ratio of 0.36 (and 0.25). 

The network is initialized with random phase matrix $H$ with value continuously ranging from 0 and 1. Young's modulus and Poisson ratio for both materials are also randomly initialized from 0 to 0.5. For optimizer, we choose truncated Newton's method, and constrain Young's modulus to between 0 to 0.5 TPa and Poisson ration between 0 to 0.5. The termination criterion of the optimizer is set to whenever the gradient is less than 1e-9. Line search is used to determine the optimum step size.

\begin{figure}
  \centering
    \includegraphics[width=1.0\linewidth]{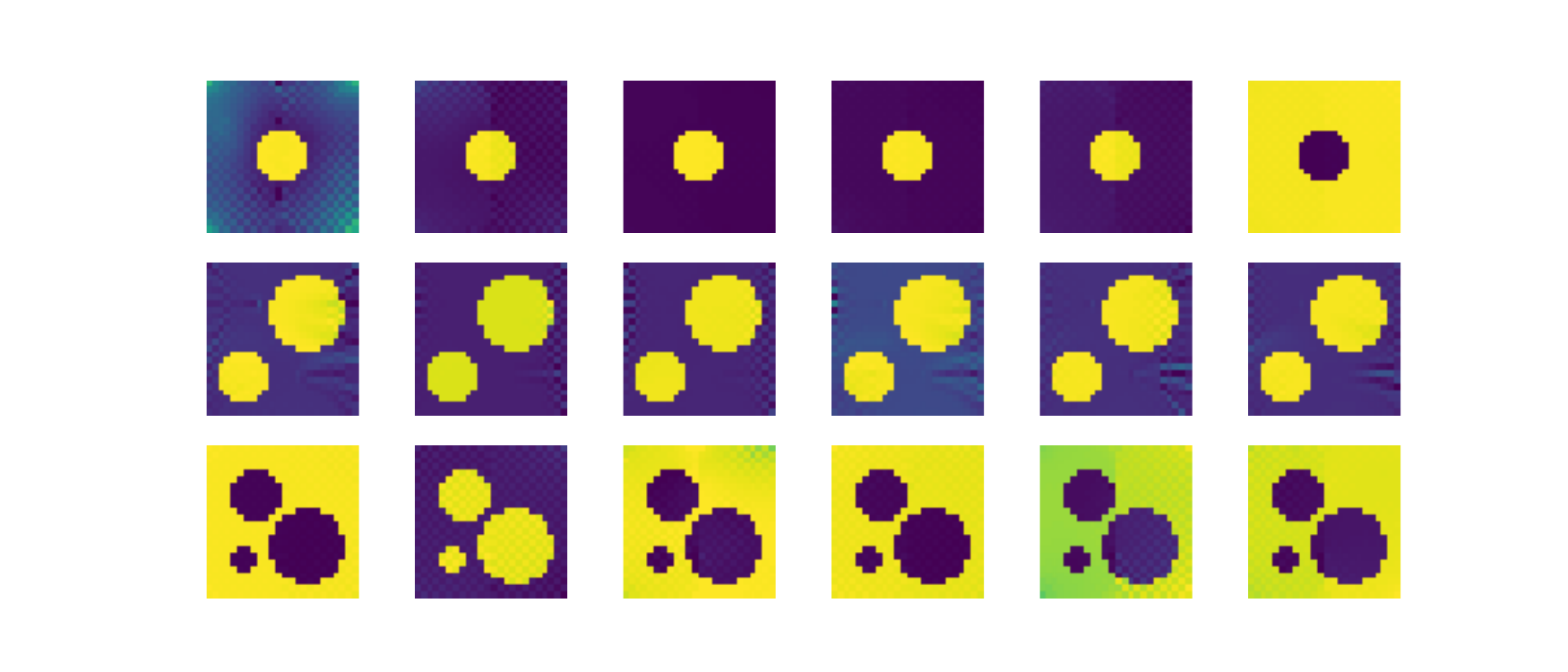}
    \caption{Visualization of the Estimated Phase Obtained from Joint Optimization of Material Phase and Property. Different columns are obtained from different random initialization. Yellow and blue color denotes material phase 0 and 1 respectively. The predicted material property corresponding to these material phases are listed in Tab.~\ref{tab:joint}. }
\label{fig:joint_opt}
\end{figure}

The learned material phase and property are shown in Fig.~\ref{fig:joint_opt} and Tab.~\ref{tab:joint}, respectively. Yellow and blue color in Fig.~\ref{fig:joint_opt} represents different material phases.
It is interesting to see that the color for inclusions and exclusion in Fig.~\ref{fig:joint_opt} can shuffle (For example, the last image in the first row of .~\ref{fig:joint_opt} has reversed color.). In the meantime, their estimated material properties for two phases are also shuffled in Tab.~\ref{tab:joint}. 
This is because shuffling the phase and material property together actually corresponds to exactly the same physics problem. In other words, there can be equally good solutions to the optimization problem Eq.\ref{eq:est_mask_rho}, and the original physics problem \ref{problem:est_joint} does not have a unique solution.

The estimation for Young's modulus is very accurate. The average error rate is 0.3\% and 2.5\% for different materials. The error on the Poisson ratio is larger, especially on two inclusions. This can be improved by an extra round of post-processing: Initialize the network again with binarized predicted material phase image, and solve Problem \ref{problem:est_rho} again to focus on learning the material property. 

\begin{table}%
\centering
\caption{Learned material properties with random initializations. The reference material property is (0.241  TPa, 0.36) and (0.2  TPa, 0.2) for both phases. inc1 to inc3 denotes different number of inclusions in the material, which can be seen from Fig.~\ref{fig:joint_opt}.}
\label{tab:joint}
\begin{tabular}{|c|c|cc|cc|}
\hline
\multicolumn{2}{|c|}{~} & \multicolumn{2}{c|}{Young's Modulus ( TPa) } &\multicolumn{2}{c|}{Poisson Ratio} \\
\hline
\hline
\multicolumn{2}{|c|}{\backslashbox{Init.}{Est.}}  & Material 0  & Material 1  & Material 0  & Material 1   \\
\hline
\parbox[t]{2mm}{\multirow{6}{*}{\rotatebox[origin=c]{90}{inc1}}} 
&case1 & 0.241    &0.197      &0.362    & 0.241      \\
&case2 & 0.241    &0.191      &0.362    & 0.212      \\
&case3 & 0.241    &0.189      &0.361    & 0.203     \\
&case4 & 0.241    &0.191      &0.361    & 0.214     \\
&case5 & 0.241    &0.192      &0.362    & 0.215      \\
&case6 & 0.190    &0.241      &0.205    & 0.360      \\
\hline
\parbox[t]{2mm}{\multirow{6}{*}{\rotatebox[origin=c]{90}{inc2}}} 
&case1 & 0.241    &0.204    &0.275      & 0.294      \\
&case2 & 0.241    &0.203    &0.274      & 0.295      \\
&case3 & 0.241    &0.204    &0.275      & 0.294     \\
&case4 & 0.241    &0.203    &0.274      & 0.294     \\
&case5 & 0.241    &0.204    &0.275      & 0.294      \\
&case6 & 0.241    &0.203    &0.275      & 0.294      \\
\hline
\parbox[t]{2mm}{\multirow{6}{*}{\rotatebox[origin=c]{90}{inc3}}} 
&case1 &0.198      & 0.241     & 0.241      &0.363    \\
&case2 &0.242       & 0.196    & 0.364     &0.234    \\
&case3 &0.200      & 0.242   & 0.250      &0.364    \\
&case4 &0.196       & 0.244    & 0.234    &0.371    \\
&case5 &0.193       & 0.234    & 0.221     &0.373    \\
&case6 &0.197        & 0.244   & 0.239     &0.372    \\
\hline
\end{tabular}
\end{table}

\subsection{Verification of response prediction}
\label{subsec:verification_inference}
In this subsection, we demonstrate that the response prediction for multi-physics and multi-phase is converging w.r.t. network depth. From Sec.~\ref{subsec:verification_multiphysics} and Sec.~\ref{subsec:verification_multiphase} we have seen that FEA-Net is capable to learn either physically meaningful filters or physics parameters robustly and accurately. Based on this, we investigate the convergence of the inference architecture of FEA-Net by assuming that the convolutional kernel has been learned correctly.

For multi-physics problem,  we use material property at: $E=0.212 TPa$, $\nu=0.288$, $\kappa=16 W/(m\cdot K)$, and $\alpha=1.2e^{-5} /^{\circ} C$ to generate the reference response. For multi-phase problem, the second material property is set at $E=0.23 TPa$ and $\nu=0.275$.
On the other hand, based on these physics parameters, the reference network filters can be obtained based on Eq.~\ref{eq:w_therm} to Eq.~\ref{eq:coupling} for thermoelasticity , and Eq.~\ref{eq:w_elast_xx}, Eq.~\ref{eq:w_elast_xy} for bi-phase elasticity.

\begin{figure}
  \centering
    \includegraphics[width=0.9\linewidth]{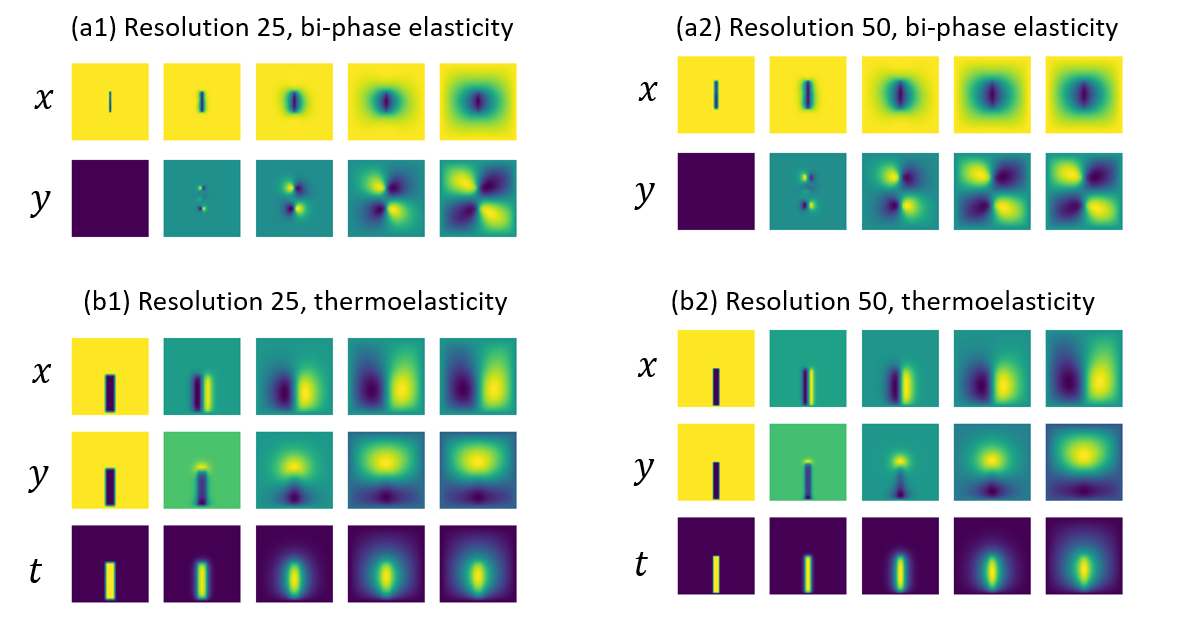}
    \caption{Visualization of FEA-Net inference output. Different rows correspond to x- and y- directional response. From left to right columns are the network input and output at 10th, 100th, 500th, and 4000th layers.}
\label{fig:inference_convergence}
\end{figure}

Since only convolution operation is involved in FEA-Net, it is able to handle inputs of different resolutions. Thus, we also test the inference performance with loading image of resolution 25 and 50. The predicted response at different network layers is visualized in Fig.~\ref{fig:inference_convergence}a and Fig.~\ref{fig:inference_convergence}b for thermoelasticity and bi-phase elasticity respectively. 
It can be seen that they are all converging well, and there is no visual difference between FEA-Net prediction at 4000 layers and ground truth for these cases.


\begin{figure}
  \centering
    \includegraphics[width=0.5\linewidth]{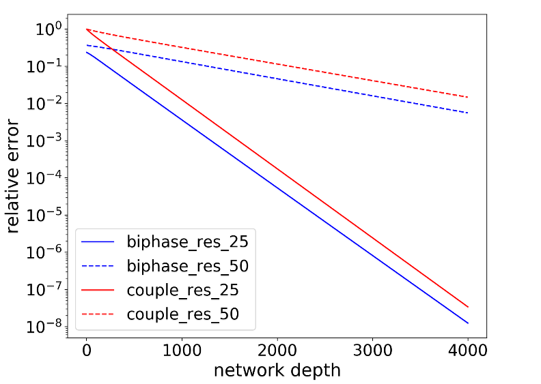}
    \caption{Error convergence for bi-phase elasticity and homogeneous thermoelasticity coupling under resolution of 25 and 50. }
\label{fig:inf_conv}
\end{figure}

We further visualize the convergence of the network prediction error w.r.t. the network depth in Fig.~\ref{fig:inf_conv}.  Interestingly, the loading image at a lower resolution is converging much faster than its higher resolution counterpart. It is conceivable that predicting the response at a higher resolution is a harder task.
We expect improvements can be made in at least three aspects: (1) Use larger filter size, which covers a larger region and has a larger receptive field. This is analogous to using higher-order finite elements as discussed in Theorem 1.
(2) Form FEA convolution block with several layers of FEA convolution layers, and use it to replace the current single FEA convolution. Stacking several convolutions together also leads to a larger receptive field.
(3) Building better network architectures based on more advanced solvers like multi-grid. The multi-grid network tends to converge much faster, as it computes the response at different resolutions.

{\color{black}
The computation time for inference network depends on its depth. As a reference, the forward pass of a single loading image with resolution 50-by-50 over a single inference block takes 0.08 ms for the multi-physics problem. 
With growing DOFs, it will take increasing number inference blocks to achieve the same level of accuracy if the inference network architecture is unchanged. But the number of layers needed can be reduced with better inference network architectures (e.g., multi-grid).
}

\subsection{Comparison with deep neural networks}
\label{sec:compare_fcn}
In this section, we compare FEA-Net with data-driven approaches in predicting homogeneous elasticity problems. Because FCN can handle image input of different resolution, it is used as the benchmark for comparison. 
As the simplest example, we train both networks on single-phase elasticity and ask it to make a prediction when a new loading is applied.

\subsubsection{Dataset and network setup}
\label{subsec:dataset}
\begin{figure}
  \centering
    \includegraphics[width=0.95\linewidth]{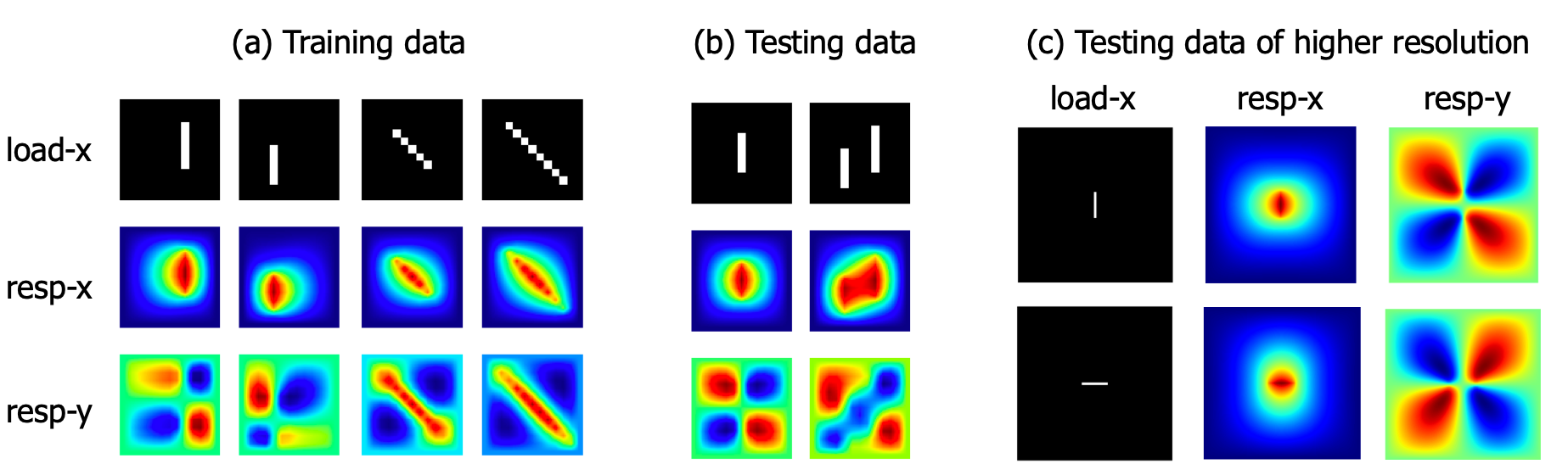}
    \hspace{0.3cm}
    \caption{Visualization of training and testing data. (a): Training set at a resolution of 12 by 12.
    (b): Testing data at resolution 12 by 12.
    (c): Testing data at resolution 60 by 60. 
    Loading is applied uniformly along the x-direction only, with locations visualized as the white regions in the plot.}
\label{fig:single_phase_data}
\end{figure}

To compare the data efficiency, our training set only includes 4 loading and response image pairs obtained from numerical simulation under different loading conditions. As shown in Fig.~\ref{fig:single_phase_data}a, the white lines in the first row are the locations where a uniform x-directional force is applied. The second and third rows are the displacement along with x and y directions respectively. Different columns correspond to different training data pair. The material used to generate these data pairs has an elasticity modulus of 0.2  TPa and Poisson's ratio of 0.25.
To thoroughly investigate the generalization capability of different models, we created four different testing cases. These testing set are are shown in Fig.~\ref{fig:single_phase_data}b and Fig.~\ref{fig:single_phase_data}c, which is composed of different loading conditions and image resolutions. 
Learning on such few amount of data can be a very challenging task for purely data-driven approaches like FCN, but can be handled by proposed FEA-Net.

\begin{figure}%
  \centering
    \includegraphics[width=0.99\linewidth]{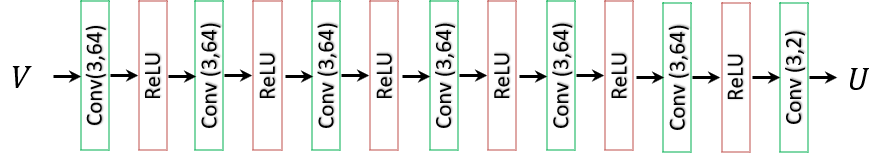}
    \caption{Architecture of the baseline FCN model. Conv(m,n) denotes a convolutional layer with filter size m-by-m and n output channels. This network has 7 convolutional layers and a forward pas has a total of 20k convolutional operations.}
\label{fig:fcn_baseline}
\end{figure}

The benchmark FCN model takes in the system loading image and outputs the predicted response image. As shown in Fig.~\ref{fig:fcn_baseline}, it has 7 layers: The first layer has 2 input channels and 64 output channels, the middle 5 layers have 64 input and 64 output channels, and the output layer has 64 input channels and 2 output channels. The filter size is kept as 3x3, which is the same as FEA-Net. ReLU activation is applied after every layer except the last one. Under such setting, the network contains over 4k filters and 186k trainable variables. The training objective of FCN is to minimize the predicted response with the reference system response. We build this network in Tensorflow and train it with Adam optimizer \citep{kingma2014adam}. 

The training architecture of FEA-Net is still one layer, with response images as input and loading image as output. The inference architecture of FEA-Net we used has 5k layers, each layer has 4 filters, which also leads to a total of 20k convolution operations.

\subsubsection{Experiment results}
\label{subsec:singlephase_analysis}
\begin{figure}%
  \centering
    \includegraphics[width=0.9
    \linewidth]{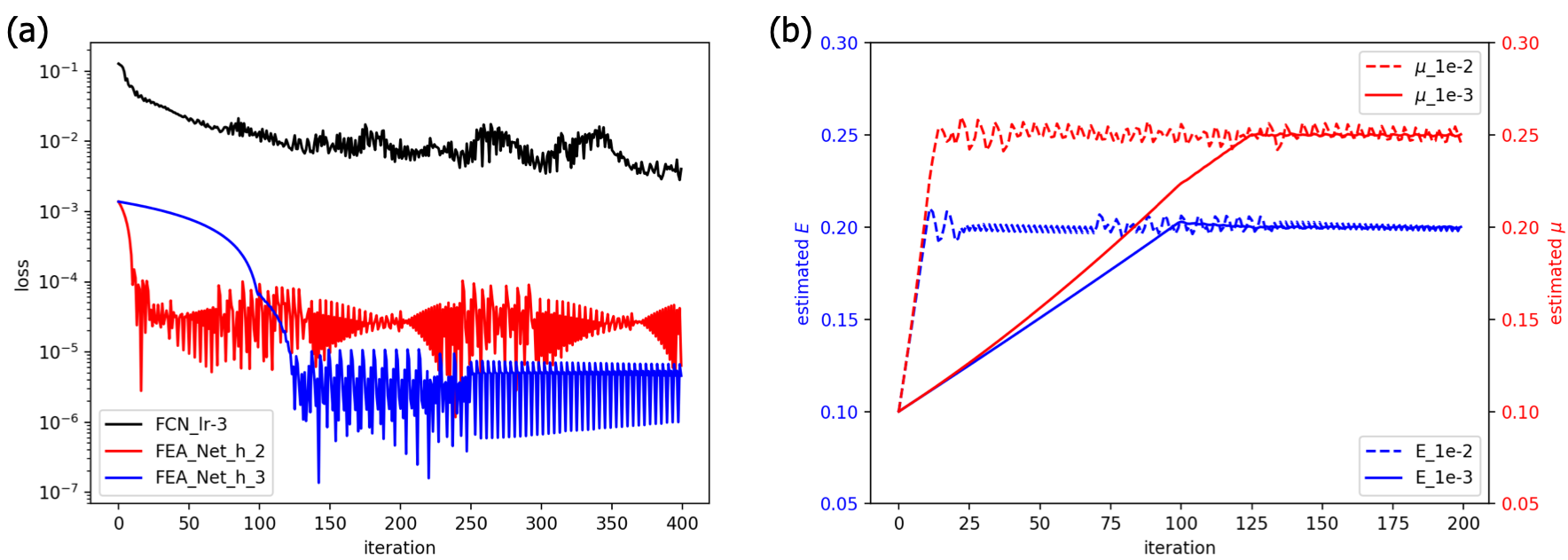}
    \caption{(a) Convergence of the training loss for FCN with learning rate $10^{-3}$ and FEA-Net with learning rate $10^{-2}$ and $10^{-3}$. (b) Convergence of the network training in material property estimation with random initialization. Reference value is $E=0.2$ and $\nu=0.25$. (The value of $E$ is scaled by $1e-12$)}
\label{fig:single_phase_training_convergence}
\end{figure}

We train both FEA-Net and FCN with Adam optimizer till converge. The convergence of their loss for the first 400 iterations is shown in Fig.~\ref{fig:single_phase_training_convergence}a. Note that the magnitude of the loss is not directly comparable, since FEA-Net is defined on the difference between reference loading and predicted loading while FCN is defined on the difference between reference response and network predicted response. However, while FCN is still not converging at 400 iterations with a learning rate of $10^{-3}$, FEA-Net is converging around 150 steps under the same learning rate. And if we further increase the learning rate for FEA-Net to $10^{-2}$, it is able to converge within 20 steps. And similar to training traditional neural networks, there is also a trade-off in learning rate. With larger learning rate the algorithm will learn faster, while able to learn more stable at a smaller learning rate.

Another major difference is that, while the filters from traditional convolutional neural networks like FCN is not interpretable, proposed FEA-Net is designed to have physics knowledge embedded and we can infer the physics parameters from its filters. As shown in Fig.~\ref{fig:single_phase_training_convergence}b, FEA-Net successfully learns the correct physics parameters under different learning rate. Similar to the convergence of the loss value, there is also a trade-off between estimation accuracy and learning speed.

\begin{figure}
  \centering
    \includegraphics[width=0.9\linewidth]{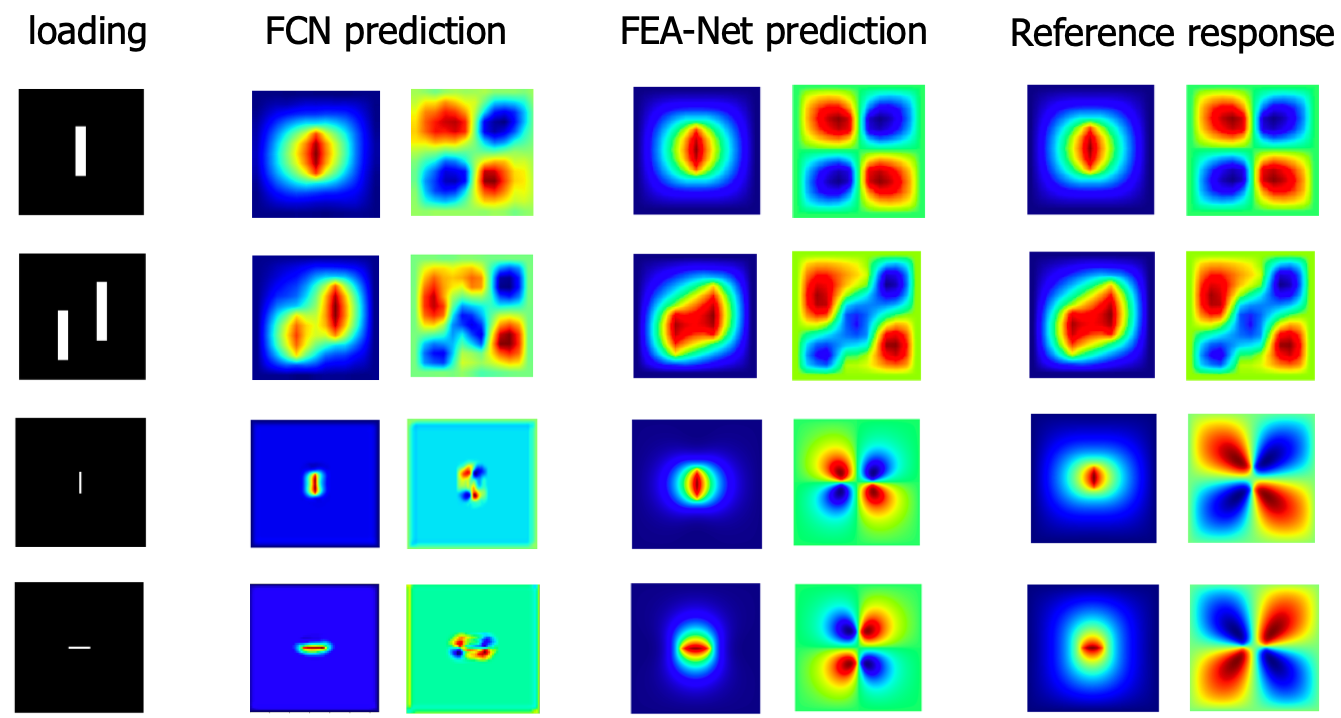}
    \caption{Comparison of the network prediction from FCN and FEA-Net. Top to bottom rows corresponds to different test loading cases and the predictions. Loading is applied uniformly along the x-direction only. Two columns in each section correspond to x and y directional components.}
\label{fig:single_phase_prediction}
\end{figure}

Once the network has been trained, we can use it to predict the response image given a new testing loading image. A visual comparison between FCN and FEA-Net is shown in the second and third columns in Fig.~\ref{fig:single_phase_prediction}. It can be seen that FCN can predict the first testing case well, which shows that our FCN model is reasonable and its training is successful.
However, FCN is not able to make correct predictions for all other testing cases which are getting more and more different than the loading images. Although it seems that FCN is getting the correct trend, its prediction is still far away from the ground truth.  
Such a result suggests that there is a big problem with the generalizability of FCN.
We also visualize the prediction of FEA-Net with 5000 layers in the fourth and fifth columns in Fig.~\ref{fig:single_phase_prediction}. It can be seen that there is almost no visual difference between the network predictions and the ground truth. That is to say, the proposed model can generalize well with limited training data. 

\subsection{Comparison with FEA}
\label{sec:compare_fea}

We derive the memory consumption for FEA-Net and traditional FEA in solving bi-phase elasticity problem. In this subsection, we use $n$ to denote the resolution of the loading image. {\color{black} Since classical FEA uses double precision by default, we also assume that double precision is used in the proposed method for fair comparison.}
The memory consumption for traditional FEA is estimated by considering only the storage for the sparse representation of the stiffness matrix $K$ and the loading vector. The loading will be stored with double-precision floating numbers (8 bytes), which costs $8*2*n^2$ byte memory as each node has two loading components. And since the bandwidth of the stiffness matrix for 2D elasticity is roughly 18, it requires $18*8*n^2$ byte to store the values of stiffness matrix with sparse representation. Besides, an additional $18*4*2*n^2$ byte memory is needed to store the row and column index of the sparse matrix with unsigned long int. Summing them together, FEA requires $304n^2$ byte memory in total. 
On the other hand, the convolutional filter is shared across all layers in FEA-Net and only very few amount of memory is needed.  We estimate the cost for FEA-Net as the storage needed for the filter $W$ (which is $4*9*8$ byte), loading image with 2 channels ($8*2*n^2$ byte), and the phase image ($8*n^2$ byte). Summing them together, FEA-Net requires $24n^2+288$ byte memory in theory, which is 12.7 times less than FEA. 
The benefit in storage-saving can be more significant for 3D problems or problems involving in multi-physics where the bandwidth of the stiffness matrix is larger. For example, the bandwidth of the stiffness matrix is increased to 81 for 3D thermal problems, the memory savings can be 1/55 with the proposed method. A list of the comparison is given in Tab.~\ref{tab:data_usage} on the memory consumption of different situations.

\begin{table}
\centering
\caption{Comparison on Memory Usage. $n$ is the resolution.}
\begin{tabular}{C{1.3cm}|C{1.2cm}C{1.4cm}|C{1.45cm}C{1.6cm}|C{1.45cm}C{1.6cm}|C{1.45cm}C{1.6cm}}\hline
  \multicolumn{1}{c|}{\multirow{2}{*}{Problem}} & \multicolumn{2}{c|}{thermal} & \multicolumn{2}{c|}{elasticity}& \multicolumn{2}{c|}{bi-phase elasticity}& \multicolumn{2}{c}{thermoelasticity}\\ \cline{2-9}
  & 2D & 3D & 2D & 3D & 2D & 3D & 2D & 3D\\ 
  \hline
  \multirow{1}{*}{FEA} & $152n^2$ & $440n^3$& $304n^2$ & $1320n^3$& $304n^2$ & $1304n^3$ &$456n^2$ &$1760n^3$  \\ 
  \multirow{1}{*}{Proposed} &$8n^2+72$ &$8n^3+216$ & $16n^2+288$ & $24n^3+1944$ & $24n^2+288$ & $32n^3+1944$ &$24n^2+648$ &$32n^3+3456$ \\ 
  \hline
  \multirow{1}{*}{Ratio}  &19.0 &55.0 & 19.0 & 55.0 & 12.7 & 40.8 & 19.0 & 55.0 \\ 
  \hline
\end{tabular}
\label{tab:data_usage}
\end{table}

\begin{figure}
  \centering
    \includegraphics[width=0.9\linewidth]{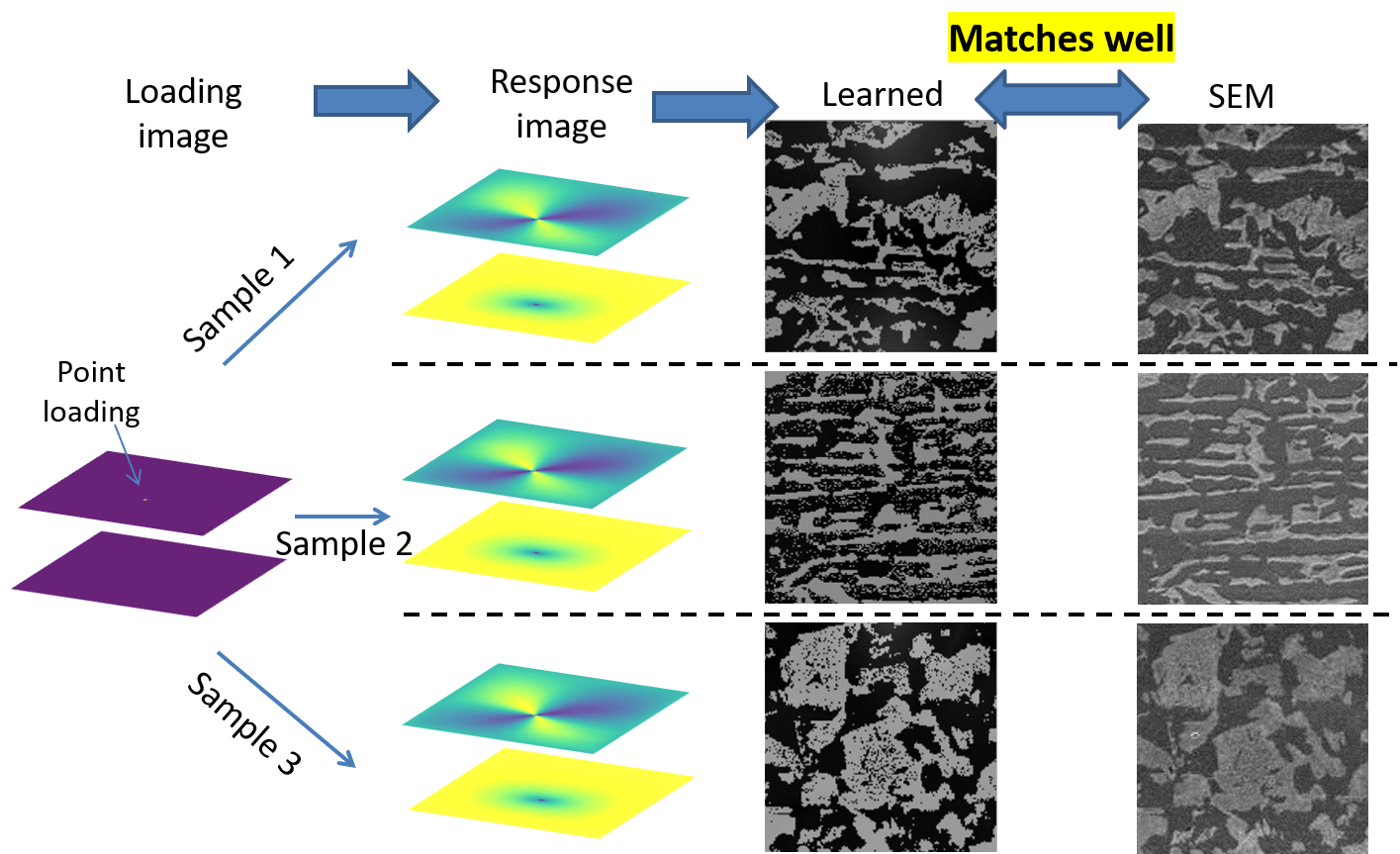}
    \caption{The estimated phase of the pipeline samples. The first column shows the loading image, with the response image on different samples are shown in the second column. The third and fourth column shows the learned material phase and phase obtained from SEM scan.}
\label{fig:micro_result}
\end{figure}

We further use a real-world problem to compare the memory consumption of the proposed method with FEA. 
We have a pipeline system installed between the year 1949 to 1961, and wish to analyze its micro-structure to monitor its health conditions. As in \citep{dahire2018pipeline}, it is known to us that the pipeline is composed of two phases, ferrite and pearlite. Since we know the material property and wish to learn its phase from loading/ response data, this fails into Problem \ref{problem:est_phase}. 
The loading and response image pairs we used has a resolution of 150, which is shown in the first and second columns in Fig.~\ref{fig:micro_result}. The learned phase images for different samples are shown in the third column of Fig.~\ref{fig:micro_result}. And the reference phase images obtained from Scanning Electron Microscope (SEM) are shown in the last column of Fig.~\ref{fig:micro_result}.  The learned material phase images and reference ones match very well with each other.

\begin{figure}%
  \centering
    \includegraphics[width=0.6\linewidth]{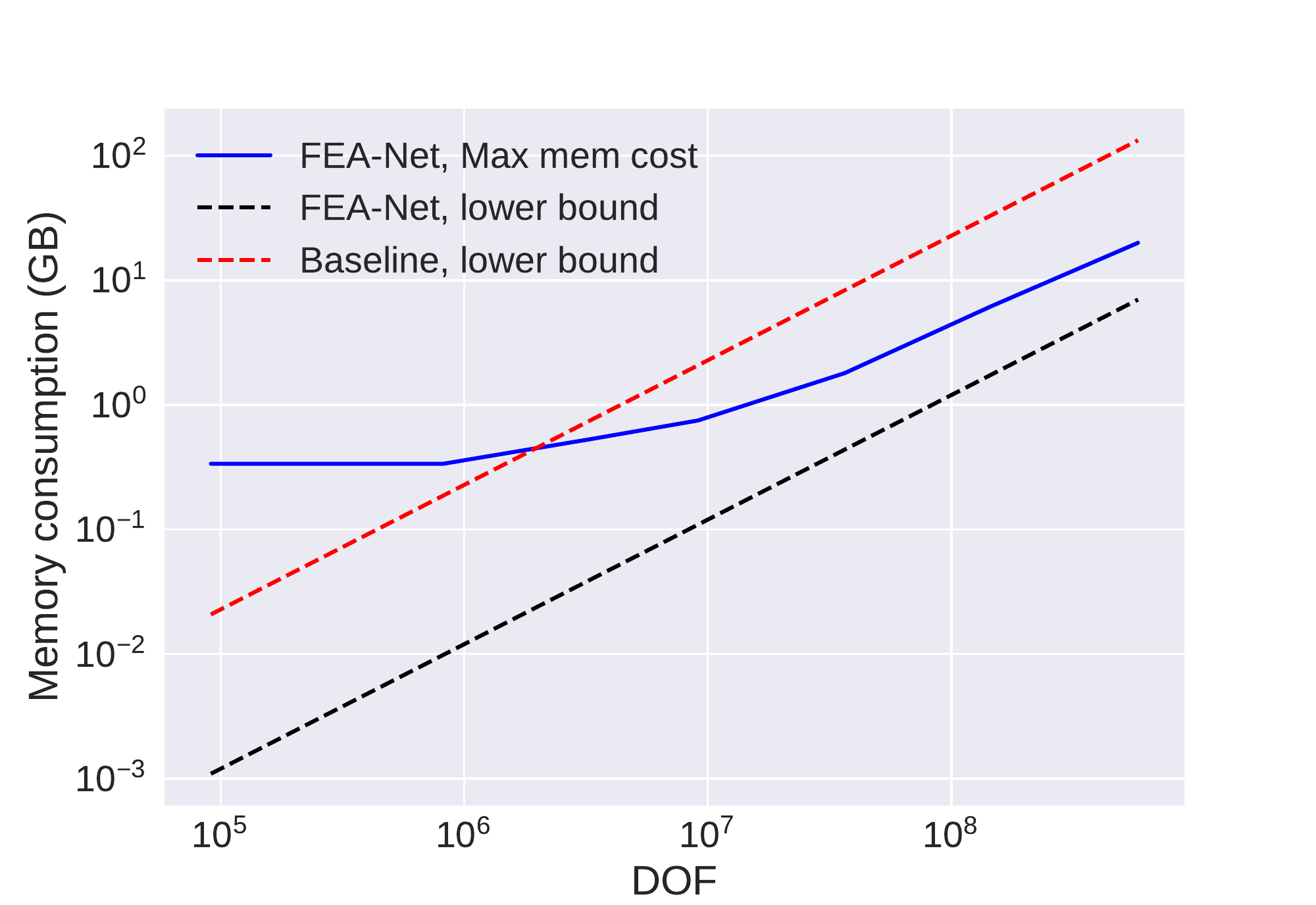}
    \caption{Comparison of Memory Cost with Standard FEA. Red and black dashed lines are the estimated lower bound of memory consumption. The blue curve is the peak memory consumption of our model obtained from the experiment.}
\label{fig:memory_cost}
\end{figure}

The peak memory consumption of the proposed method under our Tensorflow implementation is shown in Fig.~\ref{fig:memory_cost}. It can be seen that the proposed method has larger memory consumption a lower DOF, which is caused by the overhead of Tensorflow implementation. However, as DOF increases, the memory consumption of our implementation is approaching the theoretical bound. And our method will start to consume less memory than the baseline lower bound starting from 3 million DOF. 

As for the cost in computational time, if sparse matrix-vector production is used, the complexity for traditional FEA should be the same to FEA-Net. However, since convolution can be implemented very efficiently on GPU, there could be an improvement on the time consumption considering the benefit from the hardware side.

\section{Discussion and conclusion}
\label{sec:conclusion}

Motivated by the success and limitations of both data-driven models and physics-based models, we present a hybrid learning approach to predict the physics response with limited training data samples.
The proposed method is a very flexible model, which can have different physics prior knowledge added easily. 
It has good interpretability, as the network filters are designed to reflect the PDE behind the physics behavior. 
Theoretical analysis and empirical experiments have shown that the proposed method is very data efficient in learning and has good convergence at predicting both multi-physics and multi-phase problems.  


Furthermore, there are many interesting directions worth pursuing based on this study: (1) By setting the second material to have zero mechanical property, the proposed method can handle topology optimization.
(2) More efficient inference architectures can be built with more advanced solvers like multi-grid.
(3) Multi-grid solvers can be used to model material homogenization at a different scale such as in \citep{liu2019heterogeneouzation}.
(4) Extending current networks to non-linear to model the non-linearity in the material property. 
(5) Use larger convolution filters or more convolutional layers for higher efficiency. 
(6) {\color{black} Extend current model to irregular mesh by treating them as graphs and incorporate the newly developed graph convolutions (and graph convolutional neural networks) \cite{Hanocka_2019_meshcnn} to learn the stiffness matrix.}
(7) Learning on complex shapes and geometries. {\color{black}This is relatively straight forward with graph convolution operations. Another approach is to solve the problem inside a larger bounding box and then filter out the solutions that does not belong to the region of interest [25].}

\section*{Acknowledgement}
We would like to thank Dr. Yi Ren for the helpful suggestions in experiments and paper writing, Dr. Yuzhong Chen for proofreading the paper, and Haoyang Wei, Dr. Yang Yu for helping generating the experiment data.
The research reported in this paper was partially supported by funds from NASA University Leadership Initiative program (Contract No. NNX17AJ86A, PI: Yongming Liu, Project Officer: Anupa Bajwa). The support is gratefully acknowledged.

\bibliographystyle{model1-num-names}
\bibliography{sample.bib}

\begin{thebibliography}{52}
\expandafter\ifx\csname natexlab\endcsname\relax\def\natexlab#1{#1}\fi
\providecommand{\bibinfo}[2]{#2}
\ifx\xfnm\relax \def\xfnm[#1]{\unskip,\space#1}\fi
\bibitem[{Liu et~al.(2006)Liu, Stratman, and Mahadevan}]{liu2006fatigue}
\bibinfo{author}{Y.~Liu}, \bibinfo{author}{B.~Stratman},
  \bibinfo{author}{S.~Mahadevan},
\newblock \bibinfo{title}{Fatigue crack initiation life prediction of railroad
  wheels},
\newblock \bibinfo{journal}{International journal of fatigue}
  \bibinfo{volume}{28} (\bibinfo{year}{2006}) \bibinfo{pages}{747--756}.
\bibitem[{Maqsood et~al.(2004)Maqsood, Khan, and Abraham}]{maqsood2004ensemble}
\bibinfo{author}{I.~Maqsood}, \bibinfo{author}{M.~R. Khan},
  \bibinfo{author}{A.~Abraham},
\newblock \bibinfo{title}{An ensemble of neural networks for weather
  forecasting},
\newblock \bibinfo{journal}{Neural Computing \& Applications}
  \bibinfo{volume}{13} (\bibinfo{year}{2004}) \bibinfo{pages}{112--122}.
\bibitem[{Dutta(2002)}]{dutta2002geopressure}
\bibinfo{author}{N.~Dutta},
\newblock \bibinfo{title}{Geopressure prediction using seismic data: Current
  status and the road ahead},
\newblock \bibinfo{journal}{Geophysics} \bibinfo{volume}{67}
  (\bibinfo{year}{2002}) \bibinfo{pages}{2012--2041}.
\bibitem[{Krizhevsky et~al.(2012)Krizhevsky, Sutskever, and
  Hinton}]{krizhevsky2012imagenet}
\bibinfo{author}{A.~Krizhevsky}, \bibinfo{author}{I.~Sutskever},
  \bibinfo{author}{G.~E. Hinton},
\newblock \bibinfo{title}{Imagenet classification with deep convolutional
  neural networks},
\newblock in: \bibinfo{booktitle}{Advances in neural information processing
  systems}, pp. \bibinfo{pages}{1097--1105}.
\bibitem[{Amodei et~al.(2016)Amodei, Ananthanarayanan, Anubhai, Bai,
  Battenberg, Case, Casper, Catanzaro, Cheng, Chen
  et~al.}]{amodei2016deepspeech}
\bibinfo{author}{D.~Amodei}, \bibinfo{author}{S.~Ananthanarayanan},
  \bibinfo{author}{R.~Anubhai}, \bibinfo{author}{J.~Bai},
  \bibinfo{author}{E.~Battenberg}, \bibinfo{author}{C.~Case},
  \bibinfo{author}{J.~Casper}, \bibinfo{author}{B.~Catanzaro},
  \bibinfo{author}{Q.~Cheng}, \bibinfo{author}{G.~Chen}, et~al.,
\newblock \bibinfo{title}{Deep speech 2: End-to-end speech recognition in
  english and mandarin},
\newblock in: \bibinfo{booktitle}{International Conference on Machine
  Learning}, pp. \bibinfo{pages}{173--182}.
\bibitem[{Devlin et~al.(2018)Devlin, Chang, Lee, and
  Toutanova}]{devlin2018bert}
\bibinfo{author}{J.~Devlin}, \bibinfo{author}{M.-W. Chang},
  \bibinfo{author}{K.~Lee}, \bibinfo{author}{K.~Toutanova},
\newblock \bibinfo{title}{Bert: Pre-training of deep bidirectional transformers
  for language understanding},
\newblock \bibinfo{journal}{arXiv preprint arXiv:1810.04805}
  (\bibinfo{year}{2018}).
\bibitem[{Silver et~al.(2017)Silver, Schrittwieser, Simonyan, Antonoglou,
  Huang, Guez, Hubert, Baker, Lai, Bolton et~al.}]{silver2017mastering}
\bibinfo{author}{D.~Silver}, \bibinfo{author}{J.~Schrittwieser},
  \bibinfo{author}{K.~Simonyan}, \bibinfo{author}{I.~Antonoglou},
  \bibinfo{author}{A.~Huang}, \bibinfo{author}{A.~Guez},
  \bibinfo{author}{T.~Hubert}, \bibinfo{author}{L.~Baker},
  \bibinfo{author}{M.~Lai}, \bibinfo{author}{A.~Bolton}, et~al.,
\newblock \bibinfo{title}{Mastering the game of go without human knowledge},
\newblock \bibinfo{journal}{Nature} \bibinfo{volume}{550}
  (\bibinfo{year}{2017}) \bibinfo{pages}{354}.
\bibitem[{Sheikholeslami et~al.(2019)Sheikholeslami, Gerdroodbary, Moradi,
  Shafee, and Li}]{sheikholeslami2019nanofluid}
\bibinfo{author}{M.~Sheikholeslami}, \bibinfo{author}{M.~B. Gerdroodbary},
  \bibinfo{author}{R.~Moradi}, \bibinfo{author}{A.~Shafee},
  \bibinfo{author}{Z.~Li},
\newblock \bibinfo{title}{Application of neural network for estimation of heat
  transfer treatment of al2o3-h2o nanofluid through a channel},
\newblock \bibinfo{journal}{Computer Methods in Applied Mechanics and
  Engineering} \bibinfo{volume}{344} (\bibinfo{year}{2019})
  \bibinfo{pages}{1--12}.
\bibitem[{Tompson et~al.(2017)Tompson, Schlachter, Sprechmann, and
  Perlin}]{tompson2016fluid}
\bibinfo{author}{J.~Tompson}, \bibinfo{author}{K.~Schlachter},
  \bibinfo{author}{P.~Sprechmann}, \bibinfo{author}{K.~Perlin},
\newblock \bibinfo{title}{Accelerating eulerian fluid simulation with
  convolutional networks},
\newblock in: \bibinfo{booktitle}{Proceedings of the 34th International
  Conference on Machine Learning-Volume 70}, \bibinfo{organization}{JMLR. org},
  pp. \bibinfo{pages}{3424--3433}.
\bibitem[{Chu and Thuerey(2017)}]{chu2017smoke}
\bibinfo{author}{M.~Chu}, \bibinfo{author}{N.~Thuerey},
\newblock \bibinfo{title}{Data-driven synthesis of smoke flows with cnn-based
  feature descriptors},
\newblock \bibinfo{journal}{ACM Transactions on Graphics (TOG)}
  \bibinfo{volume}{36} (\bibinfo{year}{2017}) \bibinfo{pages}{69}.
\bibitem[{Wang et~al.(2019)Wang, Zhang, Sun, and Wu}]{wang2019load_path}
\bibinfo{author}{Q.~Wang}, \bibinfo{author}{G.~Zhang},
  \bibinfo{author}{C.~Sun}, \bibinfo{author}{N.~Wu},
\newblock \bibinfo{title}{High efficient load paths analysis with u* index
  generated by deep learning},
\newblock \bibinfo{journal}{Computer Methods in Applied Mechanics and
  Engineering} \bibinfo{volume}{344} (\bibinfo{year}{2019})
  \bibinfo{pages}{499--511}.
\bibitem[{Finol et~al.(2018)Finol, Lu, Mahadevan, and
  Srivastava}]{finol2018eigen}
\bibinfo{author}{D.~Finol}, \bibinfo{author}{Y.~Lu},
  \bibinfo{author}{V.~Mahadevan}, \bibinfo{author}{A.~Srivastava},
\newblock \bibinfo{title}{Deep convolutional neural networks for eigenvalue
  problems in mechanics},
\newblock \bibinfo{journal}{International Journal for Numerical Methods in
  Engineering}  (\bibinfo{year}{2018}).
\bibitem[{Sosnovik and Oseledets(2019)}]{sosnovik2017neural}
\bibinfo{author}{I.~Sosnovik}, \bibinfo{author}{I.~Oseledets},
\newblock \bibinfo{title}{Neural networks for topology optimization},
\newblock \bibinfo{journal}{Russian Journal of Numerical Analysis and
  Mathematical Modelling} \bibinfo{volume}{34} (\bibinfo{year}{2019})
  \bibinfo{pages}{215--223}.
\bibitem[{Cang et~al.(2019)Cang, Yao, and Ren}]{cang2018cad}
\bibinfo{author}{R.~Cang}, \bibinfo{author}{H.~Yao}, \bibinfo{author}{Y.~Ren},
\newblock \bibinfo{title}{One-shot generation of near-optimal topology through
  theory-driven machine learning},
\newblock \bibinfo{journal}{Computer-Aided Design} \bibinfo{volume}{109}
  (\bibinfo{year}{2019}) \bibinfo{pages}{12--21}.
\bibitem[{Bouman et~al.(2013)Bouman, Xiao, Battaglia, and
  Freeman}]{bouman2013estimating}
\bibinfo{author}{K.~L. Bouman}, \bibinfo{author}{B.~Xiao},
  \bibinfo{author}{P.~Battaglia}, \bibinfo{author}{W.~T. Freeman},
\newblock \bibinfo{title}{Estimating the material properties of fabric from
  video},
\newblock in: \bibinfo{booktitle}{Proceedings of the IEEE international
  conference on computer vision}, pp. \bibinfo{pages}{1984--1991}.
\bibitem[{Li et~al.(2019)Li, Liu, Cui, Luo, Li, and
  Zhuang}]{li2019predicting_shale}
\bibinfo{author}{X.~Li}, \bibinfo{author}{Z.~Liu}, \bibinfo{author}{S.~Cui},
  \bibinfo{author}{C.~Luo}, \bibinfo{author}{C.~Li},
  \bibinfo{author}{Z.~Zhuang},
\newblock \bibinfo{title}{Predicting the effective mechanical property of
  heterogeneous materials by image based modeling and deep learning},
\newblock \bibinfo{journal}{Computer Methods in Applied Mechanics and
  Engineering} \bibinfo{volume}{347} (\bibinfo{year}{2019})
  \bibinfo{pages}{735--753}.
\bibitem[{Bessa et~al.(2017)Bessa, Bostanabad, Liu, Hu, Apley, Brinson, Chen,
  and Liu}]{bessa2017framework}
\bibinfo{author}{M.~Bessa}, \bibinfo{author}{R.~Bostanabad},
  \bibinfo{author}{Z.~Liu}, \bibinfo{author}{A.~Hu}, \bibinfo{author}{D.~W.
  Apley}, \bibinfo{author}{C.~Brinson}, \bibinfo{author}{W.~Chen},
  \bibinfo{author}{W.~K. Liu},
\newblock \bibinfo{title}{A framework for data-driven analysis of materials
  under uncertainty: Countering the curse of dimensionality},
\newblock \bibinfo{journal}{Computer Methods in Applied Mechanics and
  Engineering} \bibinfo{volume}{320} (\bibinfo{year}{2017})
  \bibinfo{pages}{633--667}.
\bibitem[{Cang et~al.(2018)Cang, Li, Yao, Jiao, and Ren}]{cang2018improving}
\bibinfo{author}{R.~Cang}, \bibinfo{author}{H.~Li}, \bibinfo{author}{H.~Yao},
  \bibinfo{author}{Y.~Jiao}, \bibinfo{author}{Y.~Ren},
\newblock \bibinfo{title}{Improving direct physical properties prediction of
  heterogeneous materials from imaging data via convolutional neural network
  and a morphology-aware generative model},
\newblock \bibinfo{journal}{Computational Materials Science}
  \bibinfo{volume}{150} (\bibinfo{year}{2018}) \bibinfo{pages}{212--221}.
\bibitem[{Zhao et~al.(2019)Zhao, Yan, Chen, Mao, Wang, and
  Gao}]{zhao2019DL_SHM}
\bibinfo{author}{R.~Zhao}, \bibinfo{author}{R.~Yan}, \bibinfo{author}{Z.~Chen},
  \bibinfo{author}{K.~Mao}, \bibinfo{author}{P.~Wang}, \bibinfo{author}{R.~X.
  Gao},
\newblock \bibinfo{title}{Deep learning and its applications to machine health
  monitoring},
\newblock \bibinfo{journal}{Mechanical Systems and Signal Processing}
  \bibinfo{volume}{115} (\bibinfo{year}{2019}) \bibinfo{pages}{213--237}.
\bibitem[{Yao et~al.(2019)Yao, Wen, Ren, Wu, and Ji}]{yao2019sensorcalibration}
\bibinfo{author}{H.~Yao}, \bibinfo{author}{J.~Wen}, \bibinfo{author}{Y.~Ren},
  \bibinfo{author}{B.~Wu}, \bibinfo{author}{Z.~Ji},
\newblock \bibinfo{title}{Low-cost measurement of industrial shock signals via
  deep learning calibration},
\newblock in: \bibinfo{booktitle}{ICASSP 2019-2019 IEEE International
  Conference on Acoustics, Speech and Signal Processing (ICASSP)},
  \bibinfo{organization}{IEEE}, pp. \bibinfo{pages}{2892--2896}.
\bibitem[{Wang and Yao(2019)}]{wang2019FSL}
\bibinfo{author}{Y.~Wang}, \bibinfo{author}{Q.~Yao},
\newblock \bibinfo{title}{Few-shot learning: A survey},
\newblock \bibinfo{journal}{arXiv preprint arXiv:1904.05046}
  (\bibinfo{year}{2019}).
\bibitem[{Abadi et~al.(2016)Abadi, Barham, Chen, Chen, Davis, Dean, Devin,
  Ghemawat, Irving, Isard et~al.}]{abadi2016tensorflow}
\bibinfo{author}{M.~Abadi}, \bibinfo{author}{P.~Barham},
  \bibinfo{author}{J.~Chen}, \bibinfo{author}{Z.~Chen},
  \bibinfo{author}{A.~Davis}, \bibinfo{author}{J.~Dean},
  \bibinfo{author}{M.~Devin}, \bibinfo{author}{S.~Ghemawat},
  \bibinfo{author}{G.~Irving}, \bibinfo{author}{M.~Isard}, et~al.,
\newblock \bibinfo{title}{Tensorflow: A system for large-scale machine
  learning},
\newblock in: \bibinfo{booktitle}{12th $\{$USENIX$\}$ Symposium on Operating
  Systems Design and Implementation ($\{$OSDI$\}$ 16)}, pp.
  \bibinfo{pages}{265--283}.
\bibitem[{Paszke et~al.(2017)Paszke, Gross, Chintala, Chanan, Yang, DeVito,
  Lin, Desmaison, Antiga, and Lerer}]{paszke2017pytorch}
\bibinfo{author}{A.~Paszke}, \bibinfo{author}{S.~Gross},
  \bibinfo{author}{S.~Chintala}, \bibinfo{author}{G.~Chanan},
  \bibinfo{author}{E.~Yang}, \bibinfo{author}{Z.~DeVito},
  \bibinfo{author}{Z.~Lin}, \bibinfo{author}{A.~Desmaison},
  \bibinfo{author}{L.~Antiga}, \bibinfo{author}{A.~Lerer},
\newblock \bibinfo{title}{Automatic differentiation in pytorch}
  (\bibinfo{year}{2017}).
\bibitem[{Kawaguchi(2016)}]{kawaguchi201no_local_minima}
\bibinfo{author}{K.~Kawaguchi},
\newblock \bibinfo{title}{Deep learning without poor local minima},
\newblock in: \bibinfo{booktitle}{Advances in neural information processing
  systems}, pp. \bibinfo{pages}{586--594}.
\bibitem[{Hsieh et~al.(2018)Hsieh, Zhao, Eismann, Mirabella, and
  Ermon}]{hsieh2018ICLR_PDE}
\bibinfo{author}{J.-T. Hsieh}, \bibinfo{author}{S.~Zhao},
  \bibinfo{author}{S.~Eismann}, \bibinfo{author}{L.~Mirabella},
  \bibinfo{author}{S.~Ermon},
\newblock \bibinfo{title}{Learning neural pde solvers with convergence
  guarantees}  (\bibinfo{year}{2018}).
\bibitem[{Hornik(1991)}]{hornik1991universal}
\bibinfo{author}{K.~Hornik},
\newblock \bibinfo{title}{Approximation capabilities of multilayer feedforward
  networks},
\newblock \bibinfo{journal}{Neural networks} \bibinfo{volume}{4}
  (\bibinfo{year}{1991}) \bibinfo{pages}{251--257}.
\bibitem[{LeCun et~al.(1989)LeCun, Boser, Denker, Henderson, Howard, Hubbard,
  and Jackel}]{lecun1989lenet}
\bibinfo{author}{Y.~LeCun}, \bibinfo{author}{B.~Boser}, \bibinfo{author}{J.~S.
  Denker}, \bibinfo{author}{D.~Henderson}, \bibinfo{author}{R.~E. Howard},
  \bibinfo{author}{W.~Hubbard}, \bibinfo{author}{L.~D. Jackel},
\newblock \bibinfo{title}{Backpropagation applied to handwritten zip code
  recognition},
\newblock \bibinfo{journal}{Neural computation} \bibinfo{volume}{1}
  (\bibinfo{year}{1989}) \bibinfo{pages}{541--551}.
\bibitem[{Girshick et~al.(2014)Girshick, Donahue, Darrell, and
  Malik}]{girshick2014rcnn}
\bibinfo{author}{R.~Girshick}, \bibinfo{author}{J.~Donahue},
  \bibinfo{author}{T.~Darrell}, \bibinfo{author}{J.~Malik},
\newblock \bibinfo{title}{Rich feature hierarchies for accurate object
  detection and semantic segmentation},
\newblock in: \bibinfo{booktitle}{Proceedings of the IEEE conference on
  computer vision and pattern recognition}, pp. \bibinfo{pages}{580--587}.
\bibitem[{Ren et~al.(2015)Ren, He, Girshick, and Sun}]{ren2015frcnn}
\bibinfo{author}{S.~Ren}, \bibinfo{author}{K.~He},
  \bibinfo{author}{R.~Girshick}, \bibinfo{author}{J.~Sun},
\newblock \bibinfo{title}{Faster r-cnn: Towards real-time object detection with
  region proposal networks},
\newblock in: \bibinfo{booktitle}{Advances in neural information processing
  systems}, pp. \bibinfo{pages}{91--99}.
\bibitem[{Radford et~al.(2015)Radford, Metz, and Chintala}]{radford2015dcgan}
\bibinfo{author}{A.~Radford}, \bibinfo{author}{L.~Metz},
  \bibinfo{author}{S.~Chintala},
\newblock \bibinfo{title}{Unsupervised representation learning with deep
  convolutional generative adversarial networks},
\newblock \bibinfo{journal}{arXiv preprint arXiv:1511.06434}
  (\bibinfo{year}{2015}).
\bibitem[{Isola et~al.(2017)Isola, Zhu, Zhou, and Efros}]{isola2017image2image}
\bibinfo{author}{P.~Isola}, \bibinfo{author}{J.-Y. Zhu},
  \bibinfo{author}{T.~Zhou}, \bibinfo{author}{A.~A. Efros},
\newblock \bibinfo{title}{Image-to-image translation with conditional
  adversarial networks},
\newblock in: \bibinfo{booktitle}{Proceedings of the IEEE conference on
  computer vision and pattern recognition}, pp. \bibinfo{pages}{1125--1134}.
\bibitem[{Long et~al.(2015)Long, Shelhamer, and Darrell}]{long2015fcn}
\bibinfo{author}{J.~Long}, \bibinfo{author}{E.~Shelhamer},
  \bibinfo{author}{T.~Darrell},
\newblock \bibinfo{title}{Fully convolutional networks for semantic
  segmentation},
\newblock in: \bibinfo{booktitle}{Proceedings of the IEEE conference on
  computer vision and pattern recognition}, pp. \bibinfo{pages}{3431--3440}.
\bibitem[{Ronneberger et~al.(2015)Ronneberger, Fischer, and
  Brox}]{ronneberger2015unet}
\bibinfo{author}{O.~Ronneberger}, \bibinfo{author}{P.~Fischer},
  \bibinfo{author}{T.~Brox},
\newblock \bibinfo{title}{U-net: Convolutional networks for biomedical image
  segmentation},
\newblock in: \bibinfo{booktitle}{International Conference on Medical image
  computing and computer-assisted intervention},
  \bibinfo{organization}{Springer}, pp. \bibinfo{pages}{234--241}.
\bibitem[{He et~al.(2016)He, Zhang, Ren, and Sun}]{he2016resnet}
\bibinfo{author}{K.~He}, \bibinfo{author}{X.~Zhang}, \bibinfo{author}{S.~Ren},
  \bibinfo{author}{J.~Sun},
\newblock \bibinfo{title}{Deep residual learning for image recognition},
\newblock in: \bibinfo{booktitle}{Proceedings of the IEEE conference on
  computer vision and pattern recognition}, pp. \bibinfo{pages}{770--778}.
\bibitem[{Huang et~al.(2017)Huang, Liu, Weinberger, and van~der
  Maaten}]{huang2017densely}
\bibinfo{author}{G.~Huang}, \bibinfo{author}{Z.~Liu}, \bibinfo{author}{K.~Q.
  Weinberger}, \bibinfo{author}{L.~van~der Maaten},
\newblock \bibinfo{title}{Densely connected convolutional networks},
\newblock in: \bibinfo{booktitle}{Proceedings of the IEEE conference on
  computer vision and pattern recognition}, volume~\bibinfo{volume}{1},
  p.~\bibinfo{pages}{3}.
\bibitem[{Xie et~al.(2017)Xie, Girshick, Doll{\'a}r, Tu, and
  He}]{xie2017resnext}
\bibinfo{author}{S.~Xie}, \bibinfo{author}{R.~Girshick},
  \bibinfo{author}{P.~Doll{\'a}r}, \bibinfo{author}{Z.~Tu},
  \bibinfo{author}{K.~He},
\newblock \bibinfo{title}{Aggregated residual transformations for deep neural
  networks},
\newblock in: \bibinfo{booktitle}{Proceedings of the IEEE conference on
  computer vision and pattern recognition}, pp. \bibinfo{pages}{1492--1500}.
\bibitem[{Fei-Fei et~al.(2006)Fei-Fei, Fergus, and Perona}]{fei2006oneshot}
\bibinfo{author}{L.~Fei-Fei}, \bibinfo{author}{R.~Fergus},
  \bibinfo{author}{P.~Perona},
\newblock \bibinfo{title}{One-shot learning of object categories},
\newblock \bibinfo{journal}{IEEE transactions on pattern analysis and machine
  intelligence} \bibinfo{volume}{28} (\bibinfo{year}{2006})
  \bibinfo{pages}{594--611}.
\bibitem[{Santoro et~al.(2016)Santoro, Bartunov, Botvinick, Wierstra, and
  Lillicrap}]{santoro2016meta}
\bibinfo{author}{A.~Santoro}, \bibinfo{author}{S.~Bartunov},
  \bibinfo{author}{M.~Botvinick}, \bibinfo{author}{D.~Wierstra},
  \bibinfo{author}{T.~Lillicrap},
\newblock \bibinfo{title}{Meta-learning with memory-augmented neural networks},
\newblock in: \bibinfo{booktitle}{International conference on machine
  learning}, pp. \bibinfo{pages}{1842--1850}.
\bibitem[{Hughes(2012)}]{hughes2012fea}
\bibinfo{author}{T.~J. Hughes}, \bibinfo{title}{The finite element method:
  linear static and dynamic finite element analysis},
  \bibinfo{publisher}{Courier Corporation}, \bibinfo{year}{2012}.
\bibitem[{Yang and Mittal(2014)}]{yang2014SRJ}
\bibinfo{author}{X.~I. Yang}, \bibinfo{author}{R.~Mittal},
\newblock \bibinfo{title}{Acceleration of the jacobi iterative method by
  factors exceeding 100 using scheduled relaxation},
\newblock \bibinfo{journal}{Journal of Computational Physics}
  \bibinfo{volume}{274} (\bibinfo{year}{2014}) \bibinfo{pages}{695--708}.
\bibitem[{Li(2000)}]{li2000gfs_cmame}
\bibinfo{author}{S.~Li},
\newblock \bibinfo{title}{Global flexibility simulation and element stiffness
  simulation in finite element analysis with neural network},
\newblock \bibinfo{journal}{Computer Methods in Applied Mechanics and
  Engineering} \bibinfo{volume}{186} (\bibinfo{year}{2000})
  \bibinfo{pages}{101--108}.
\bibitem[{Oishi and Yagawa(2017)}]{oishi2017enhance_fea}
\bibinfo{author}{A.~Oishi}, \bibinfo{author}{G.~Yagawa},
\newblock \bibinfo{title}{Computational mechanics enhanced by deep learning},
\newblock \bibinfo{journal}{Computer Methods in Applied Mechanics and
  Engineering} \bibinfo{volume}{327} (\bibinfo{year}{2017})
  \bibinfo{pages}{327--351}.
\bibitem[{Capuano and Rimoli(2019)}]{capuano2019smart_FEA}
\bibinfo{author}{G.~Capuano}, \bibinfo{author}{J.~J. Rimoli},
\newblock \bibinfo{title}{Smart finite elements: A novel machine learning
  application},
\newblock \bibinfo{journal}{Computer Methods in Applied Mechanics and
  Engineering} \bibinfo{volume}{345} (\bibinfo{year}{2019})
  \bibinfo{pages}{363--381}.
\bibitem[{Yu et~al.(2018)Yu, Yao, and Liu}]{yu2018physics}
\bibinfo{author}{Y.~Yu}, \bibinfo{author}{H.~Yao}, \bibinfo{author}{Y.~Liu},
\newblock \bibinfo{title}{Physics-based learning for aircraft dynamics
  simulation},
\newblock in: \bibinfo{booktitle}{PHM Society Conference},
  volume~\bibinfo{volume}{10}.
\bibitem[{Lu et~al.(2018)Lu, Zhong, Li, and Dong}]{lu2017beyond}
\bibinfo{author}{Y.~Lu}, \bibinfo{author}{A.~Zhong}, \bibinfo{author}{Q.~Li},
  \bibinfo{author}{B.~Dong},
\newblock \bibinfo{title}{Beyond finite layer neural networks: Bridging deep
  architectures and numerical differential equations},
\newblock in: \bibinfo{editor}{J.~Dy}, \bibinfo{editor}{A.~Krause} (Eds.),
  \bibinfo{booktitle}{Proceedings of the 35th International Conference on
  Machine Learning}, volume~\bibinfo{volume}{80} of
  \textit{\bibinfo{series}{Proceedings of Machine Learning Research}},
  \bibinfo{publisher}{PMLR}, \bibinfo{address}{Stockholmsmässan, Stockholm
  Sweden}, \bibinfo{year}{2018}, pp. \bibinfo{pages}{3282--3291}.
\bibitem[{Long et~al.(2018)Long, Lu, Ma, and Dong}]{long2018pde}
\bibinfo{author}{Z.~Long}, \bibinfo{author}{Y.~Lu}, \bibinfo{author}{X.~Ma},
  \bibinfo{author}{B.~Dong},
\newblock \bibinfo{title}{Pde-net: Learning pdes from data},
\newblock in: \bibinfo{booktitle}{Proceedings of the 35th International
  Conference on Machine Learning (ICML 2018)}.
\bibitem[{Yao et~al.(2019)Yao, Ren, and Liu}]{yao2019fea}
\bibinfo{author}{H.~Yao}, \bibinfo{author}{Y.~Ren}, \bibinfo{author}{Y.~Liu},
\newblock \bibinfo{title}{Fea-net: A deep convolutional neural network with
  physicsprior for efficient data driven pde learning},
\newblock in: \bibinfo{booktitle}{AIAA Scitech 2019 Forum}, p.
  \bibinfo{pages}{0680}.
\bibitem[{Rumelhart et~al.(1988)Rumelhart, Hinton, Williams
  et~al.}]{rumelhart1988back_propogation}
\bibinfo{author}{D.~E. Rumelhart}, \bibinfo{author}{G.~E. Hinton},
  \bibinfo{author}{R.~J. Williams}, et~al.,
\newblock \bibinfo{title}{Learning representations by back-propagating errors},
\newblock \bibinfo{journal}{Cognitive modeling} \bibinfo{volume}{5}
  (\bibinfo{year}{1988}) \bibinfo{pages}{1}.
\bibitem[{Kingma and Ba(2014)}]{kingma2014adam}
\bibinfo{author}{D.~P. Kingma}, \bibinfo{author}{J.~Ba},
\newblock \bibinfo{title}{Adam: A method for stochastic optimization},
\newblock \bibinfo{journal}{arXiv preprint arXiv:1412.6980}
  (\bibinfo{year}{2014}).
\bibitem[{Dahire et~al.(2018)Dahire, Tahir, Jiao, and Liu}]{dahire2018pipeline}
\bibinfo{author}{S.~Dahire}, \bibinfo{author}{F.~Tahir},
  \bibinfo{author}{Y.~Jiao}, \bibinfo{author}{Y.~Liu},
\newblock \bibinfo{title}{Bayesian network inference for probabilistic strength
  estimation of aging pipeline systems},
\newblock \bibinfo{journal}{International Journal of Pressure Vessels and
  Piping} \bibinfo{volume}{162} (\bibinfo{year}{2018}) \bibinfo{pages}{30--39}.
\bibitem[{Liu et~al.(2019)Liu, Wu, and Koishi}]{liu2019heterogeneouzation}
\bibinfo{author}{Z.~Liu}, \bibinfo{author}{C.~Wu}, \bibinfo{author}{M.~Koishi},
\newblock \bibinfo{title}{A deep material network for multiscale topology
  learning and accelerated nonlinear modeling of heterogeneous materials},
\newblock \bibinfo{journal}{Computer Methods in Applied Mechanics and
  Engineering} \bibinfo{volume}{345} (\bibinfo{year}{2019})
  \bibinfo{pages}{1138--1168}.
\bibitem[{Hanocka et~al.(2019)Hanocka, Hertz, Fish, Giryes, Fleishman, and
  Cohen-Or}]{Hanocka_2019_meshcnn}
\bibinfo{author}{R.~Hanocka}, \bibinfo{author}{A.~Hertz},
  \bibinfo{author}{N.~Fish}, \bibinfo{author}{R.~Giryes},
  \bibinfo{author}{S.~Fleishman}, \bibinfo{author}{D.~Cohen-Or},
\newblock \bibinfo{title}{Meshcnn: A network with an edge},
\newblock \bibinfo{journal}{ACM Trans. Graph.} \bibinfo{volume}{38}
  (\bibinfo{year}{2019}) \bibinfo{pages}{90:1--90:12}.

\end{thebibliography}

\pagebreak
\appendix
\section{Analytical FEA convolutional kernels}

We derive the analytical form of the FEA convolutional kernels for different physics problems in this appendix. The geometry matrix $B$ in Eq.~\ref{eq:element_stiffness} has an expression of:
\begin{equation}
\label{eq:B_mat}
    B = L N
\end{equation}
where $L$ is differential operator.

For simplicity, we choose $\Delta$ to be the simplest linear element in Eq.~\ref{eq:element_stiffness}, which makes $N$ has the form of:   
\begin{equation}
\label{eq:N_mat}
    N = \frac{1}{4}
    \begin{bmatrix}
    (1-\xi)(1-\eta) & (1+\xi)(1-\eta) & (1+\xi)(1+\eta) & (1-\xi)(1+\eta)
    \end{bmatrix}
\end{equation}

For thermal and elasticity problems, we have:
\begin{equation}
\label{eq:B1_mat}
    B^1 = 
    \begin{bmatrix}[1.5]
    \frac{\partial N^e_1}{\partial \xi} & \frac{\partial N^e_2}{\partial \xi} & \frac{\partial N^e_3}{\partial \xi} & \frac{\partial N^e_4}{\partial \xi}\\
    \frac{\partial N^e_1}{\partial \eta} & \frac{\partial N^e_2}{\partial \eta} & \frac{\partial N^e_3}{\partial \eta} & \frac{\partial N^e_4}{\partial \eta}\\
    \end{bmatrix}\\
\end{equation}
\begin{equation}
\label{eq:B2_mat}
    B^2 = 
    \begin{bmatrix}[1.5]
    \frac{\partial N^e_1}{\partial \xi} & \frac{\partial N^e_2}{\partial \xi} & \frac{\partial N^e_3}{\partial \xi} & \frac{\partial N^e_4}{\partial \xi} & 0 & 0 & 0 & 0\\
    0 & 0 & 0 & 0 & \frac{\partial N^e_1}{\partial \eta} & \frac{\partial N^e_2}{\partial \eta} & \frac{\partial N^e_3}{\partial \eta} & \frac{\partial N^e_4}{\partial \eta}\\
    \frac{\partial N^e_1}{\partial \eta} & \frac{\partial N^e_2}{\partial \eta} & \frac{\partial N^e_3}{\partial \eta} & \frac{\partial N^e_4}{\partial \eta} & \frac{\partial N^e_1}{\partial \xi} & \frac{\partial N^e_2}{\partial \xi} & \frac{\partial N^e_3}{\partial \xi} & \frac{\partial N^e_4}{\partial \xi}
    \end{bmatrix}\\
\end{equation}
And the constitutional matrix $C$ differs for different problems.

Once we have $B$ and $C$ defined, we can compute each term of the element stiffness matrix by integrating Eq.~\ref{eq:element_stiffness}. The integral has relatively simple forms in many cases, and analytical solutions can be directly obtained. 
We use $\hat K$ to represent the element stiffness matrix in this appendix to avoid duplication in notation.

\subsection{Thermal problem}
\label{appendix:thermal_kernel}
Thermal problems are governed by Poisson equation:
\begin{equation}
    \kappa (u_{,xx} + u_{,yy} ) = v
\end{equation}
where $u$ and $v$ denotes temperature and heat flux, and $\kappa$ is the heat conductivity ratio. 

The matrix $C$ has expression:
\begin{equation}
\label{eq:c_thermo}
    C =
    \begin{bmatrix}
    \kappa & 0\\
    0 & k\\
    \end{bmatrix}
\end{equation}
By substituting Eq.~\ref{eq:B_mat} and Eq.~\ref{eq:c_thermo} into Eq.~\ref{eq:element_stiffness}, we have:
\begin{equation}
{
    \hat K = \frac{1}{16} \int_{-1}^{1} \int_{-1}^{1}   
    \begin{bmatrix}
    -(\xi-1)^2-(\eta-1)^2 & \xi^2+\eta^2-2y & \xi^2-\eta^2-2 & \xi^2+\eta^2-2\xi\\
    \xi^2+\eta^2-2\eta & -(\xi+1)^2-(\eta-1)^2 & \xi^2+\eta^2+2\xi & \xi^2+\eta^2-2\\
    \xi^2-\eta^2-2 & \xi^2+\eta^2+2\xi & -(\xi+1)^2-(\eta+1)^2 & \xi^2+\eta^2+2\eta\\
    \xi^2+\eta^2-2\xi & \xi^2+\eta^2-2 & \xi^2+\eta^2+2\eta & -(\xi-1)^2-(\eta+1)^2\\
    \end{bmatrix}
    \text{d} \xi \text{d} \eta
}
\end{equation}
This integration can be computed analytically: 
\begin{equation}
    \hat K = \frac{\kappa}{6}
    \begin{bmatrix}
    -4 & 1 & 2 & 1\\
    1 & -2 & 1 & 2\\
    2 & 1 & -4 & 1\\
    1 & 2 & 1 & -4\\
    \end{bmatrix}
\end{equation}
By further substituting to Eq.~\ref{eq:conv_kernel_expression}, we have the FEA convolutional kernel for the thermal problem:
\begin{equation}
\begin{split}
    W^{tt} &= \frac{\kappa}{3}
    \begin{bmatrix}
    1  &1  &1\\
    1 &-8 &1\\
    1 &1 &1
    \end{bmatrix}\\
\end{split}
\end{equation}

\subsection{Elasticity problem}
\label{appendix:elasticity_kernel}
Because both the loading and response for 2D elasticity have both x and y component, the FEA convolution filter $W_{} \in R^{(3,3,2,2)}$ which has 2 input channels and two output channels. 

2D plane elasticity problems are governed by the following equilibrium equation:
\begin{equation}
    C\nabla^2 u +b = 0
\end{equation}
where $u$ is the temperature and $b$ is the body force.
The matrix $C$ for plane elasticity  has expression:
\begin{equation}
\label{eq:c_elast}
    C = \frac{E}{1-\nu^2} 
    \begin{bmatrix}
    1 & \nu & 0\\
    \nu & 1 & 0\\
    0 & 0 & \frac{1-\nu}{2}
    \end{bmatrix}
\end{equation}
By substituting the constitutional matrix $C$ in Eq.~\ref{eq:c_elast} and the geometry matrix $B$ in Eq.~\ref{eq:B1_mat} into Eq.~\ref{eq:element_stiffness}, we can obtain the corresponding element stiffness matrix:
\begin{equation}
{
\hat K = 
    \frac{E}{16(1-\nu^2)}
    \begin{bmatrix}
    8-\frac{8}{3}\nu & 2\nu+2 & -\frac{4}{3}\nu-4 & 6\nu-2 & \frac{4}{3}\nu-4 & -2\nu-2 & \frac{8}{3}\nu & 2-6\nu \\
    -\frac{4}{3}\nu-4 & 2-6\nu & 8-\frac{8}{3}\nu & -2\nu-2 & \frac{8}{3}\nu & 6\nu-2 &\frac{4}{3}\nu -4 & 2\nu+2 \\
    \frac{4}{3}\nu-4 & -2\nu-2 & \frac{8}{3}\nu & 2-6\nu & 8-\frac{8}{3}\nu & 2\nu+2 & -\frac{4}{3}\nu-4 & 6\nu-2\\
    \frac{8}{3}\nu & 6\nu-2 &\frac{4}{3}\nu -4 & 2\nu+2 & -\frac{4}{3}\nu-4 & 2-6\nu & 8-\frac{8}{3}\nu & -2\nu-2\\
    2\nu+2 & 8-\frac{8}{3}\nu & 2-6\nu & \frac{8}{3}\nu & -2\nu-2 & -\frac{4}{3}\nu-4 & 6\nu-2 &\frac{4}{3}\nu -4 \\
    6\nu-2 & \frac{8}{3}\nu & -2\nu-2 & 8-\frac{8}{3}\nu & 2-6\nu & -\frac{4}{3}\nu-4 & 2\nu+2 &\frac{4}{3}\nu -4\\
    -2\nu-2 & -\frac{4}{3}\nu-4 & 6\nu-2 &\frac{4}{3}\nu -4 & 2\nu+2 & 8-\frac{8}{3}\nu & 2-6\nu & \frac{8}{3}\nu\\
    2-6\nu & -\frac{4}{3}\nu-4 & 2\nu+2 &\frac{4}{3}\nu -4 & 6\nu-2 & \frac{8}{3}\nu & -2\nu-2 & 8-\frac{8}{3}\nu\\
    \end{bmatrix}\\
}
\label{eq:elast_stiffness}
\end{equation}

where the first and second half of the rows (and columns) corresponds to x directional response (and loading). We will start by considering only the relationship between x directional loading and x directional response by extracting the entries from the upper-left section of the matrix:
\begin{equation}
\begin{split}
\hat K^{xx} = 
    \frac{E}{16(1-\nu^2)}
    \begin{bmatrix}
    8-\frac{8}{3}\nu & -\frac{4}{3}\nu-4 & \frac{4}{3}\nu-4 & \frac{8}{3}\nu \\
    -\frac{4}{3}\nu-4 & 8-\frac{8}{3}\nu & \frac{8}{3}\nu &\frac{4}{3}\nu -4 \\
    \frac{4}{3}\nu-4 & \frac{8}{3}\nu & 8-\frac{8}{3}\nu & -\frac{4}{3}\nu-4\\
    \frac{8}{3}\nu &\frac{4}{3}\nu -4 & -\frac{4}{3}\nu-4 & 8-\frac{8}{3}\nu\\
    \end{bmatrix}\\
\end{split}
\label{eq:elast_stiffness_xx}
\end{equation}
Based on Eq.~\ref{eq:conv_kernel_expression}, we can find the FEA convolutional kernel for x directional loading and response:
\begin{equation}
\label{w_elast1}
\begin{split}
    &W^{xx} = \frac{E}{4(1-\nu^2)}
    \begin{bmatrix}
        -(1-\nu/3)        & 4\nu/3       & -(1-\nu/3)\\
        -2(1+\nu/3)       & 8(1-\nu/3)    & -2(1+\nu/3)\\
        -(1-\nu/3)        & 4\nu/3       & -(1-\nu/3)
    \end{bmatrix}\\
\end{split}
\end{equation}

Similarly, the relationship between x directional loading and y directional response $W_{xy}$ can be obtained from the upper right section of Eq.~\ref{eq:elast_stiffness}, the relationship between y directional loading and y directional response $W^{yy}$ can be obtained lower right section, and the relationship between y directional loading and x directional response $W^{yx}$ can be obtained from the lower-left section. Since similar approach is used, we skip the repeated derivation and give their expressions directly:
\begin{equation}
\label{w_elast2}
\begin{split}
    &W^{xy} = W^{yx} = \frac{E}{8(1-\nu)}
    \begin{bmatrix}
        1       & 0       & -1\\
        0       & 0       & 0\\
        -1       & 0       & 1
    \end{bmatrix}\\
    &W^{yy} = \frac{E}{4(1-\nu^2)}
    \begin{bmatrix}
        -(1-\nu/3)        & -2(1+\nu/3)       & -(1-\nu/3)\\
        4\nu/3            & 8(1-\nu/3)       & 4\nu/3 \\
        -(1-\nu/3)        &-2(1+\nu/3)        & -(1-\nu/3)
    \end{bmatrix}\\
    &W^{yy} = 
    \frac{E}{4(1-\nu^2)}
    \Bigg(
    \begin{bmatrix}
        -1        & -2       & -1\\
        0            & 8       & 0 \\
        -1        &-2        & -1
    \end{bmatrix}
    +
    \frac{\nu}{3}
    \begin{bmatrix}
        1        & -2       & 1\\
        4        & -8       & 4 \\
        1        & -2        & 1
    \end{bmatrix}
    \Bigg)
\end{split}
\end{equation}

\subsection{Thermoelasticity problem}
\label{appendix:coupling}
The equilibrium equation of the coupled thermoelastic problems can be expressed as the following tensor form:
\begin{equation}
    \frac{1}{2}E_{ijkl}(u_{k,lj}+u_{l,kj})-E_{ijkl}\alpha\delta_{kl}\Delta T_{,j} + b_i =0 
\end{equation}
where $u$ is the displacement and $b$ is the external body force. $\alpha$ is the thermal expansion coefficient of the isotropic materials. $E_{ijkl}$ is the elastic tensor. By discretization, the matrix form of finite element analysis can be obtained,
\begin{equation}
\label{eq:c_couple}
    \begin{bmatrix}
    K^{u} & K^{ut}\\
    0 & K^t\\
    \end{bmatrix}
    \begin{bmatrix}
    u\\
    T\\
    \end{bmatrix}=
    \begin{bmatrix}
    F\\
    Q\\
    \end{bmatrix}
\end{equation}

The non-coupled stiffness matrix $K^{u}$ and $K^{t}$ are the same as previous ones. Only the coupling term $K^{ut}$ is shown here,
\begin{equation}
    \hat K^{ut} =\frac{\alpha  E}{16(\nu-1)} \int_{\Delta}
{
    \begin{bmatrix}
    (\xi-1)(\eta-1)^2 & -(\xi+1)(\eta-1)^2 & (\eta^2-1)(\xi+1) & -(\eta^2)(\xi-1)\\
    (\xi-1)^2(\eta-1) & -(\xi^2-1)(\eta-1) & (\xi^2-1)(\eta+1) & -(\xi-1)^2(\eta+1)\\
    -(\xi-1)(\eta-1)^2 & (\xi+1)(\eta-1)^2 & -(\eta^2-1)(\xi+1) & (\eta^2-1)(\xi-1)\\
    -(\xi^2-1)(\eta-1) & (\xi+1)^2(\eta-1) & -(\xi+1)^2(\eta+1) & (\xi^2-1)(\eta+1)\\
    (\eta^2-1)(\xi-1) & -(\eta^2-1)(\xi+1) & (\xi+1)(\eta+1)^2 & -(\xi-1)(\eta+1)^2\\
    (\xi^2-1)(\eta-1) & -(\xi+1)^2(\eta-1) & (\xi+1)^2(\eta+1) & -(\xi^2-1)(\eta+1)\\
    -(\eta^2-1)(\xi-1) & (\eta^2-1)(\xi+1) & -(\xi+1)(\eta+1)^2 & (\xi-1)(\eta+1)^2\\
    -(\xi-1)^2(\eta-1) & (\xi^2-1)(\eta-1) & -(\xi^2-1)(\eta+1) & (\xi-1)^2(\eta+1)\\
    \end{bmatrix}
}
    d\Omega
\end{equation}

After integration on [-1,1], we have:
\begin{equation}
    \begin{split}
\hat K = 
    \frac{\alpha E}{6(1 - \nu)}
        \begin{bmatrix}
 -2& -2& -1& -1 \\
 -2& -1& -1& -2\\
 2&  2&  1&  1\\
 -1& -2& -2& -1\\
 1&  1&  2&  2\\
 1&  2&  2&  1\\
-1& -1& -2& -2\\
 2&  1&  1&  2\\
        \end{bmatrix}
    \end{split}
\end{equation}
where odd and rows corresponds to x and y directional elasticity response. 
By extracting the entries  from the odd rows, the relationship between x directional loading and heat flux can be obtained:
\begin{equation}
    \begin{split}
\hat K^{xt} = 
    \frac{\alpha E}{6(1 - \nu)}
        \begin{bmatrix}
 -2& -2& -1& -1 \\
 2&  2&  1&  1\\
 1&  1&  2&  2\\
-1& -1& -2& -2\\
        \end{bmatrix}
    \end{split}
\end{equation}
Based on Eq.~\ref{eq:conv_kernel_expression}, we can find it FEA convolutional kernel:
\begin{equation}
\label{w_elast1}
\begin{split}
    &W^{xt} =     \frac{\alpha E}{6(1 - \nu)}
    \begin{bmatrix}
        -1        & 0       & 1  \\
        -4       & 0    & 4\\
        -1        & 0       & 1
    \end{bmatrix}\\
\end{split}
\end{equation}

Similarly, the FEA convolutional kernel for the relationship can be obtained as:
\begin{equation}
\label{w_elast1}
\begin{split}
    &W^{xt} =     \frac{\alpha E}{6(1 - \nu)}
    \begin{bmatrix}
        1        & 4       & 1\\
        0       & 0    & 0\\
        -1        & -4       & -1
    \end{bmatrix}\\
\end{split}
\end{equation}

\section{Forward inference related proof}
\label{appendix:inference_convergence}

By using $u^*$ (and $\hat u$) to denote known and unknown values respectively, Eq.~\ref{eq:fea_eq} can be written as:

\begin{equation}
\label{appendix:eq_inf_1}
\left[
    \begin{array}{c;{2pt/2pt}c}
        K_{00} & K_{01} \\ \hdashline[2pt/2pt]
        K_{10} & K_{11}
    \end{array}
\right]
\left[
    \begin{array}{c}
        \hat u \\ \hdashline[2pt/2pt]
        u^*
    \end{array}
\right]
=\left[
    \begin{array}{c}
        v^* \\ \hdashline[2pt/2pt]
        \hat v
    \end{array}
\right]
\end{equation}

The boundary condition operator $\mathcal{B}$ is defined as:
\begin{equation}
    \mathcal{B}
    \left[
    \begin{array}{c}
    u_1 \\ \hdashline[2pt/2pt]
    u_2
    \end{array}
    \right]
    =
    \left[
    \begin{array}{c}
    u_1 \\ \hdashline[2pt/2pt]
    u^*
    \end{array}
    \right]
\end{equation}
which makes:
\begin{equation}
    \mathcal{B} 
\Bigg(
\left[
    \begin{array}{c;{2pt/2pt}c}
        K_{00} & K_{01} \\ \hdashline[2pt/2pt]
        K_{10} & K_{11}
    \end{array}
\right]
\left[
    \begin{array}{c}
        u_1 \\ \hdashline[2pt/2pt]
        u_2
    \end{array}
\right]
\Bigg)
    =
\left[
    \begin{array}{c;{2pt/2pt}c}
        K_{00} & K_{01} \\ \hdashline[2pt/2pt]
        0 & I
    \end{array}
\right]
\left[
    \begin{array}{c}
        u_1 \\ \hdashline[2pt/2pt]
        u^*
    \end{array}
\right]
\end{equation}
This is equivalent to solving:
\begin{equation}
\label{appendix:eq_inf}
\left[
    \begin{array}{c;{2pt/2pt}c}
        K_{00} & K_{01} \\ \hdashline[2pt/2pt]
        0 & I
    \end{array}
\right]
\left[
    \begin{array}{c}
        \hat u \\ \hdashline[2pt/2pt]
        u^*
    \end{array}
\right]
=\left[
    \begin{array}{c}
        v^* \\ \hdashline[2pt/2pt]
        u^*
    \end{array}
\right]
\end{equation}
It is obvious that Eq.~\ref{appendix:eq_inf_1} and Eq.~\ref{appendix:eq_inf} actually have the same solution.
Thus, FEA-Net will have exactly the same convergence as the numerical solver to its corresponding FEA problem.

\section{Learning with multi-physics related proof}
\label{appendix:data_corelation}

The objective of training FEA-Net on linear physics boils down to learning the filter $W$ given observed ($V$, $U$) pair:
\begin{equation}
\label{eq:app_regr}
\begin{split}
    \argmin_{W} \left\Vert W \circledast U - V \right\Vert _2^2 \\
\end{split}
\end{equation}
Since convolution operation is a linear operation, Eq.~\ref{eq:app_regr} is essentially a linear regression problem.  
For thermoelasticity, it can be decomposed into three different learning problems:
\begin{subequations}
\begin{equation}
\label{eq:est_with_vx}
    \argmin_{W^{xx}, W^{xy}, W^{xt}}  
    \left\Vert
    W^{xx} \circledast U^x + W^{xy} \circledast U^y + W^{xt} \circledast U^t - V^x
    \right\Vert _2^2
\end{equation}
\begin{equation}
    \argmin_{W^{yx}, W^{yy}, W^{yt}}  
    \left\Vert
    W^{yx} \circledast U^x + W^{yy} \circledast U^y + W^{yt} \circledast U^t - V^y  
    \right\Vert _2^2
\end{equation}
\begin{equation}
    \argmin_{W^{tx}, W^{ty}, W^{tt}}  
    \left\Vert
    W^{tx} \circledast U^x + W^{ty} \circledast U^y + W^{tt} \circledast U^t - V^t    
    \right\Vert _2^2
\end{equation}
\end{subequations}

Consider optimizing Eq.~\ref{eq:est_with_vx} for example, this optimization problem is equivalent to finding the least mean square solution of:
\begin{equation}
\begin{split}
    W^{xx} \circledast U^x + W^{xy} \circledast U^y + W^{xt} \circledast U^t &= V^x\\
\end{split}
\end{equation}
which can actually be re-organized into a matrix form:
\begin{equation}
\label{eq:lms}
    \textbf{U} \cdot \overrightarrow w = \overrightarrow v
\end{equation}
where $\textbf{U} \in R^{(n^2,27)}$ and  $\overrightarrow w \in R^{(27,1)}$ are in the form of:
\begin{subequations}
\begin{equation}
\label{eq:a_matrix_expression}
    \textbf{U} = [ \textbf{U}^x, \textbf{U}^y, \textbf{U}^t]
\end{equation}
\begin{equation}
    \overrightarrow w = [\overrightarrow {w^x}, \overrightarrow {w^y}, \overrightarrow {w^t}] ^T
\end{equation}
\end{subequations}
and their components have an expression of:
\begin{subequations}
\begin{equation}
\label{eq:x_flatten}
    \overrightarrow {w^x} = [W^{xx}_{11}, W^{xx}_{12}, W^{xx}_{13}, W^{xx}_{21}, W^{xx}_{22}, W^{xx}_{23}, W^{xx}_{31}, W^{xx}_{32}, W^{xx}_{33}]
\end{equation}
\begin{equation}
    \textbf{U}^x = [\textbf{U}^x_{i-1,j-1},\textbf{U}^x_{i-1,j},\textbf{U}^x_{i+1,j},\textbf{U}^x_{i,j-1},\textbf{U}^x_{i,j},\textbf{U}^x_{i,j+1},\textbf{U}^x_{i+1,j-1},\textbf{U}^x_{i+1,j},\textbf{U}^x_{i+1,j+1}]
\end{equation}
\end{subequations}{}
There is a total of 27 variables to be learned from Eq.~\ref{eq:lms}. There are two conditions to ensure the problem is well defined:
(1) The number of rows is larger or equal to 27. This means that we need to have the image resolution at least 6-by-6.
(2) The coefficient matrix $\textbf{U}$ is column-wise full rank.



\begin{lemma}
\label{lemma:linear_dependency}
If different loading channels are linearly dependent, $\textbf{U}$ matrix will not be row-wise full rank.
\end{lemma}
\begin{proof}
We start by assuming there exists such linear dependence:
\begin{equation}
    V^x = c_1 V^y + c_2 V^t
\end{equation}
Substituting it into the relationship between loading images and response images in Theorem 1:
\begin{equation}
\begin{split}
     W^{xx} \circledast U^x + W^{xy} \circledast U^y + W^{xt} \circledast U^t = c_1 \big( W^{yx} \circledast U^y + W^{yy} \circledast U^y + W^{yt} \circledast U^t \big) \\
     + c_2 \big(  W^{tx} \circledast U^x + W^{ty} \circledast U^y + W^{tt} \circledast U^t \big)
\end{split}
\end{equation}
after simplification we have:
\begin{equation}
\begin{split}
     (W^{xx} - c_1 W^{xy} - c_2 W^{xt} ) \circledast U^x 
     + (W^{yx} - c_1 W^{yy} - c_2 W^{yt} ) \circledast U^y\\
     + (W^{tx} - c_1 W^{ty} - c_2 W^{tt} ) \circledast U^t = 0
\end{split}
\end{equation}
which can be further re-organize into matrix form:
\begin{equation}
    \textbf{U} \cdot \overrightarrow c = 0
\end{equation}
where:
\begin{equation}
c = [\overrightarrow{w^x}-c_1\overrightarrow{w^y} -c_2 \overrightarrow{w^t}, \overrightarrow{w^x}-c_1\overrightarrow{w^y} -c_2 \overrightarrow{w^t}, \overrightarrow{w^x}-c_1\overrightarrow{w^y} -c_2 \overrightarrow{w^t}]^t
\end{equation}
Thus, matrix $\textbf{U}$ has column-wise correlation and is not of column-wise full rank.
\end{proof}

For thermoelasticity specifically, we have $W^{xt}=W^{yt}=0$ and the condition for rank deficiency in Lemma \ref{lemma:linear_dependency} can be further simplified. Since $\overrightarrow{w^t} = 0$, as long as $V^x=c_1V^y$, the system $\textbf{U}$ will have multicollinearility.
Physically, that means we can not have the loading pointing towards one direction in obtaining the training data.

\section{Learning with multi-phase related proof}
\subsection{Estimating material phase}
\label{appendix:bi-phase_phase_estimation}
From Eq.~\ref{eq:mask_conv} we can see that Eq.~\ref{eq:fea_net_h} is a linear function of $H$ if the other variables ($\rho$, $V$, $U$ are known). Furthermore,  Eq.~\ref{eq:est_mask} will become a quadratic programming problem if $L_2$ error measurement is used. Thus, the solution to Problem~\ref{problem:est_phase} is unique.
Since the governing PDE is the same everywhere in $\Omega$, the objective Eq.~\ref{eq:est_mask} also holds for any $\Phi \subset \Omega$. Thus, the material phase in $\Phi$ can be obtained from:
\begin{equation}
H^*(q) =  \argmin_{H}  h(\rho,H,V(q),U(q))\\
\end{equation}
This means that the learning of material phase can be successful for arbitrary image size.


\subsection{Estimating material property}
\label{appendix:bi-phase_rho_estimation}

It is obvious that both material phases need to get present in the phase image $H$, otherwise the other material property will not get involved in the optimization.
Now we prove the second condition on $V$. We can see that $f$ is a function that is linearly related to the FEA convolution kernel $W$. Furthermore, as can be seen from Eq.~\ref{eq:w_elast}, Eq.~\ref{eq:w_elast_couple} for homogeneous material and Eq.~\ref{eq:w_elast_xx}, Eq.~\ref{eq:w_elast_xy} for bi-phase material, the Young's modulus $E$ term can be extracted from the convolutional kernel. 
In other words, for elasticity problems, Eq.~\ref{eq:fea_net_h} is decomposible w.r.t. Young's modulus $E$:
\begin{equation}
V = h(U,H,E,\nu) = E \cdot \hat h(U,H,\nu)    
\end{equation}
If we have $V \equiv 0$, there will be two possible solutions: $E \equiv 0$ or $\hat h(U,H,\nu)  \equiv 0$. 
Thus, we need to have $V(q)$ has non-zero value(s) in order to learn the correct solution. 

Learning material property with Eq.~\ref{eq:est_rho} can be very data efficient: (1) Suppose the material is homogeneous, we have $\rho \in \mathbb{R}^2$ for elasticity problems. In this case, the optimization problem will be well defined if the number of constraints is larger or equal to 2.
(2) Suppose that there exist two phases. In this case, we have $\rho \in \mathbb{R}^4$ for elasticity problems. Thus, the resolution of $V$ needs to be larger than 2x2, and at least one element needs to be non-zero.  In either case, only a single loading-response pair can be sufficient to define the optimization problem.

\end{document}